\documentclass{LMCS}

\def\dOi{9(3:14)2013}
\lmcsheading%
{\dOi}
{1--39}
{}
{}
{Mar.~\phantom06, 2012}
{Sep.~13, 2013}
{}

\pdfoutput=1

\usepackage{enumerate,hyperref}
\usepackage{eucal}
\usepackage{colonequals}
\usepackage{diagrams}
\usepackage{stmaryrd}
\usepackage{url}
\def\digcup{{\textstyle\bigcup}}
\newarrow{Dline} =====

\newcommand{\specialcell}[2][c]{%
  \begin{tabular}[#1]{@{}c@{}}#2\end{tabular}}

\newcommand\llangle{\langle\!\langle}
\newcommand\rrangle{\rangle\!\rangle}

\begin{document}
\title{Coalgebraic Characterizations of Context-Free Languages}

\author[J.~Winter]{Joost Winter\rsuper a} 
\address{{\lsuper{a,c}}CWI, Science Park 123, 1098 XG Amsterdam, The Netherlands} \email{\{winter,janr\}@cwi.nl} 
\thanks{{\lsuper a}Supported by the NWO project CoRE: Coinductive Calculi for Regular Expressions.}

\author[M.~M.~Bonsangue]{Marcello M.~Bonsangue\rsuper b}
\address{{\lsuper b}LIACS, Niels Bohrweg 1, 2333 CA Leiden, The Netherlands} 
\email{marcello@liacs.nl}

\author[J.~Rutten]{Jan Rutten\rsuper c}
\address{\vspace{-6 pt}} 

\ACMCCS{[{\bf Theory of computation}]: Formal languages and automata
  theory---Grammars and context-free languages}

\begin{abstract}
In this article, we provide three coalgebraic characterizations of the class of context-free languages, each based on the idea of adding coalgebraic structure to an existing algebraic structure by specifying output-derivative pairs. Final coalgebra semantics then gives an interpretation function into the final coalgebra of all languages with the usual output and derivative operations. The first characterization is based on systems, where each derivative is given as a finite language over the set of nonterminals; the second characterization on systems where derivatives are given as elements of a term-algebra; and the third characterization is based on adding coalgebraic structure to a class of closed (unique) fixed point expressions. We prove equivalences between these characterizations, discuss the generalization from languages to formal power series, as well as the relationship to the generalized powerset construction.
\end{abstract}

\maketitle

\section{Introduction}

The set $\mathcal{P}(A^*)$ of all formal languages over an alphabet $A$ is a final coalgebra of the functor $\mathbb{B} \times (-)^A$, where $\mathbb{B}$ is the Boolean semiring with carrier $\{ 0, 1 \}$, uniquely defined by the defining equation $1 \lor 1 = 1$. Deterministic automata are $\mathbb{B} \times (-)^A$-coalgebras and their behaviour, in terms of language acceptance, is given by the final homomorphism into $\mathcal{P}(A^*)$. A language is \emph{regular} if it is in the image of the final homomorphism from a \emph{finite} $\mathbb{B} \times (-)^A$-coalgebra to $\mathcal{P}(A^*)$. Or, equivalently by Kleene's theorem, if it is in the image of the final homomorphism from the set of \emph{regular expressions}, which constitute a $\mathbb{B} \times (-)^A$-coalgebra by means of so-called Brzozowski derivatives.

Thus the coalgebraic picture of regular languages and regular expressions is well-understood (cf.~\cite{Ru98} for details). Moreover the picture is so elementary that it has recently been possible~\cite{Si10} to generalize it to a large class of other systems, including Mealy machines, labelled transition systems, and various probabilistic automata.

In this article, we develop a similar coalgebraic picture for \emph{context-free languages}, which form another well-known class, extending regular languages. Our focus is on \emph{context-free grammars}, one of the common definition schemes for context-free languages. (Another well-known characterization is through pushdown automata, which are treated here.)

Because the set of all languages is a final coalgebra of the functor $\mathbb{B} \times (-)^A$, we model context-free languages using coalgebras for this functor. We do so in three different but ultimately equivalent ways:
\begin{enumerate}[(1)]
\item In Section~\ref{sec:cfgcoalgebra}, we view context-free grammars in Greibach normal form as coalgebras for the functor $\mathbb{B} \times \mathcal{P}_\omega((-)^*)^A$, or equivalently, as systems of behavioural differential equations, where each derivative is given as a set of words of nonterminals. These coalgebras can then be canonically extended to $\mathbb{B} \times (-)^A$-coalgebras, yielding a mapping to the final coalgebra of all languages that coincides with the classical denotational semantics of context-free grammars. For example, in the system
\begin{center}
\begin{tabular}{ccc}
$o(x) = 1$ & $x_a = \{ xy \}$ & $x_b = \emptyset$ \\
$o(y) = 0$ & $y_a = \emptyset$ & $y_b = \{ \epsilon \}$ \\
\end{tabular}
\end{center}
$x$ is mapped onto the language $\{ a^nb^n \,|\, n \in \mathbb{N} \}$.
\item In Section~\ref{sec:langeq}, we will represent context-free languages using coalgebras for the functor $\mathbb{B} \times \mathcal{T}(-)^A$, where $\mathcal{T}(X)$ is the free term algebra containing all terminals in $A$ and nonterminals in $X$, constants $\bar{0}$ and $\bar{1}$, and closed under binary operations $+$ and $\times$. These coalgebras can be seen as syntactic systems of behavioural differential equations, again can be extended to $\mathbb{B} \times (-)^A$-coalgebras, and thus be assigned a semantics through the final coalgebra mapping.
\item In Section~\ref{sec:cfe}, we will define context-free languages and power series by means of generalized regular expressions, in which the Kleene star operation is replaced by a unique fixed point operator. The set of closed expressions again can be given a $\mathbb{B} \times (-)^A$-coalgebra structure, again yielding an operational semantics through the final homomorphism into the coalgebra of all languages.
\end{enumerate}

\noindent We show that the three coalgebraic characterizations above are equivalent in the following sense: a language is context-free if and only if it is in the image of the final homomorphism of an extension of a $\mathbb{B} \times \mathcal{P}_\omega((-)^*)^A$-coalgebra; if and only if it is in the image of the final homomorphism of an extension of a $\mathbb{B} \times \mathcal{T} (-)^A$-coalgebra; if and only if it occurs as the image under the final homomorphism of some context-free expression.

In Section~\ref{sec:generalize}, we will moreover generalize the picture to a wider class of systems, namely to Moore automata where the output alphabet possesses a commutative semiring structure. Here, we obtain a coalgebraic characterization of classes of \emph{formal power series}, which can be called context-free or \emph{(constructively) algebraic}.

In Section~\ref{sec:category}, we will cast the results from this article in a categorical framework, by showing that the constructions from Sections~\ref{sec:cfgcoalgebra} and \ref{sec:langeq} can be seen as instances of the generalized powerset construction presented in \cite{SiBoBoRu10}, and highlight some of the subtleties that arise when this is attempted.

This article is an extended and revised version of~\cite{WiBoRu11}, which has been presented at the CALCO conference in 2011. Some newer results from~\cite{BoRuWi12} have been incorporated as well. The main new contributions of the present version include a more precise and formal presentation of the results, a more comprehensive treatment of the connection to the generalized powerset construction, as well as the generalization of the results in Sections~\ref{sec:langeq} and \ref{sec:cfe} to commutative semirings.

\subsection{Related work.}

In contrast to regular languages, equality of context-free languages is known to be an undecidable property~\cite{HoMoUl06}. This may explain why not so much algebraic or coalgebraic work has been devoted to study the theory of context-free languages. The first, and only, coalgebraic treatment of context-free languages we are aware of, is presented in~\cite{HaJa05}. In this article context-free languages are described indirectly, as the result of flattening finite skeletal parsed trees. The authors study context-free grammars as coalgebras for a functor different from ours, i.e. the functor $\mathcal{P}((A + (-))^*)$.

It should be noted that, in contrast to our approach, even for finitely many non-terminal symbols, only some coalgebras for this functor are ordinary context-free grammars, as infinitely many productions must be allowed in order to obtain the set of skeletal parsed trees of finite depth via finality.

Algebraically, the starting point is Kozen's complete characterization of regular languages in terms of Kleene algebras, idempotent semirings equipped with a star-operation satisfying some fixed point equations~\cite{Ko94}. In~\cite{Le91,EsLe05}, Kleene algebras have been extended with a least fixed point operator to axiomatize fragments of the theory of context-free languages. We take a similar approach, but coalgebraic in nature and substituting the Kleene star with a unique fixed point operator. Whereas~\cite{Le91,EsLe05} are interested in providing solutions to systems of equations of the form $x = t$ using least fixed points, we look at systems of behavioural differential equations and give a semantic solution in terms of context-free languages (in Section~\ref{sec:langeq}) and syntactic solutions in terms of regular expressions with unique fixed points (in Section~\ref{sec:cfe}). Regular expressions with the Kleene star replaced by a unique right-recursive fixed point operator have been studied in~\cite{SiBoRu10,Si10} for coalgebras for a large class of functors, including $\mathbb{B} \times (-)^A$. The observation that context-free languages can be seen as solutions to systems of equations dates back to~\cite{GiRi62}.

\section{Preliminaries}

In this section, some important definitions and a few elementary results from algebra, universal coalgebra, coalgebraic automata theory, and the theory of context-free languages will be recalled. For a more extensive coalgebraic treatment of languages, automata and regular expressions, see for example \cite{Ru00,Ru98,Ja06}.

\subsection{Algebraic structures}

Recall that a \emph{monoid} $(M, \cdot, 1)$ consists of a set $M$, with a binary operation $\cdot: M \times M \to M$, and an identity element $1 \in M$, such that for all $m \in M$, $1 \cdot m = m = m \cdot 1$, and for all $m, n, p \in M$, $(m \cdot n) \cdot p = m \cdot (n \cdot p)$. Following usual conventions, the symbol $\cdot$ is often omitted. A monoid is called \emph{commutative} if for all $m$ and $n$, $mn = nm$. The set $A^*$ of words over an alphabet $A$ with the empty word (here denoted as $\epsilon$) as unit, and concatenation of words as multiplication, will, in this article, be the prime example of a monoid. This monoid is the \emph{free} monoid over the alphabet $A$.

A \emph{semiring} $(K, \cdot, +, 1, 0)$ consists of a set $K$ with two binary operations $+$ and $\cdot$, such that $(K,+, 0)$ is a commutative monoid with identity element $0$, and $(K,\cdot, 1)$ is a monoid with identity $1$, and $\cdot$ distributes over $+$, i.e. $k \cdot (l + m) = kl + km$ and $(k + l) \cdot m = km + lm$ for all $k, l, m \in K$, and moreover, for all $k \in K$, $0 \cdot k = k \cdot 0 = 0$. We call a semiring \emph{idempotent} if $k + k = k$ for all $k \in K$, and call a semiring \emph{commutative} if $(K, \cdot, 1)$ is commutative.

An important instance of a semiring is the Boolean semiring $(\mathbb{B}, \land, \lor, 1, 0)$ over the set $\mathbb{B} = \{ 0, 1 \}$, which can be uniquely characterized by the equation $1 \lor 1 = 1$. This semiring, moreover, is an initial object in the category of idempotent semirings.

Given an alphabet $A$, the set $\mathcal{P}(A^*)$ of languages over $A$, can also be assigned a semiring structure $(\mathcal{P}(A^*), \cdot, \cup, \{ \epsilon \}, \emptyset)$. Here $\emptyset$ and $\cup$ have their familiar set theoretical meaning, and $\cdot$ represents language concatenation: given languages $L, M \in \mathcal{P}(A^*)$, their product is defined by
\[
LM = \{ vw \,|\, v \in L \land w \in M \}.
\]
Closely related to this semiring is the semiring $(\mathcal{P}_\omega(A^*), \cdot, \cup, \{ \epsilon \}, \emptyset)$ of finite languages over $A$, again with $\cup$ denoting union and $\cdot$ representing language concatenation. This semiring, moreover, is the free idempotent semiring over $A$.

\subsection{Category theory}

Although for the most part (with the major exception of Section \ref{sec:category}), this article only has a light dependence on category theory, we will occasionally make use of the language of category theory. When we do this, we will assume familiar the notions of categories, (endo)functors, and natural transformations. Furthermore recall that a \emph{monad} on a category $\mathbf{C}$ consists of an endofunctor $T$ on $\mathbf{C}$, together with two natural transformations $\eta: 1_\mathbf{C} \to T$, called the \emph{unit} of the monad, and $\mu: T^2 \to T$, called the \emph{multiplication} of the monad, such that $\mu \circ \mu_T = \mu \circ T \mu$, and $\mu \circ \eta_T = 1 = \mu \circ T \eta$.

In general, given a category $\mathbf{C}$ and endofunctor $T$, a \emph{$T$-algebra} consists of an object $X$ of $\mathbf{C}$, together with a mapping $\alpha: TX \to X$. Given a monad $T$ over a category $\mathbf{C}$, an \emph{algebra for the monad $T$} is a $T$-algebra $\alpha$ such that $\alpha \circ \eta_X = 1_X$, and $\alpha \circ T(\alpha) = \alpha \circ \mu_X$. We note that, given any $X$, $\mu_X: T^2X \to TX$ is the \emph{free $T$-algebra} over $X$.

For a more comprehensive background view on monads, algebras, and category theory in general, we refer to e.g.~\cite{Aw10}.

\subsection{Coalgebraic preliminaries}

A \emph{coalgebra} for an endofunctor $F: \textbf{C} \to \textbf{C}$ on a category $\mathbf{C}$ consists of a carrier set $X$ together with a map $c: X \to FX$. The functor $F$ is usually called the \emph{type} of the coalgebra.

In this article we will be concerned with coalgebras for automata and systems of behavioural differential equations, which can all be seen as of the type $K \times T(-)^A$ for a commutative semiring $K$ and an endofunctor $T: \textbf{Set} \to \textbf{Set}$. Here $A$ is a finite set (in this context also called \emph{alphabet}), $K$ is the carrier of a commutative semiring $(K, \cdot, +, 1, 0)$, and $\times$ is the Cartesian product. For a large part of this article, we will be specifically concerned with the case where $K$ is the Boolean semiring $\mathbb{B}$.

When $T$ is the identity functor $1_\textbf{Set}$, $K \times T(-)^A$ will be the familiar functor $K \times (-)^A$ representing deterministic automata with inputs in $A$ and outputs in $K$. More generally, a coalgebra for a functor of the form $K \times T(-)^A$ can be interpreted as a system of \emph{behavioural differential equations} that for a given state $x \in X$ returns a pair $(o(x), \delta(x))$, determining the output value $o(x) \in K$ of a state $x$, and offering a structured state $\delta(x)(a) \in T(X)$ for each alphabet symbol $a \in A$. When dealing with the semiring $\mathbb{B}$, $o(x) = 1$ is equivalent to the notion that $x$ is an accepting state. Typically we will write $x_a$ for $\delta(x)(a)$, call $o(x)$ the \emph{output of $x$} and $x_a$ the \emph{$a$-derivative of $x$}, highlighting the connection between our work and Brzozowski's notion of derivatives of regular expressions \cite{Br64}.

In the case of $K \times (-)^A$-coalgebras (but not for $K \times T(-)^A$-coalgebras in general), representing deterministic automata, we can extend the notion of $a$-derivative to \emph{word derivatives} $x_w$, for $w \in A^*$, by setting $x_\epsilon = x$ for the empty word $\epsilon$ and $x_{aw} = (x_a)_w$ for $a \in A$ and $w \in A^*$.

For arbitrary $F$-coalgebras, a \emph{homomorphism} from an $F$-coalgebra $(X, c)$ to another $F$-coalgebra $(Y, d)$ is a mapping $f: C \to D$, such that $d \circ f = Ff \circ c$.

When dealing with $K \times (-)^A$-coalgebras, a function $f: X \to Y$ is a homomorphism iff $f: C \to D$ preserves outputs and next states, that is, for all $x \in X$:
\[
o(f(x)) = o(x) \quad \textrm{and} \quad f(x_a) = f(x)_a
\]

This property can easily be extended to word derivatives, as follows:
\begin{prop}
If $f: X \to Y$ is a homomorphism from a $K \times (-)^A$-coalgebra $(X, (o, \delta))$ to a $K \times (-)^A$-coalgebra $(Y, (o', \delta'))$, we have $f(x_w) = f(x)_w$ for all $x \in X$ and $w \in A^*$.
\end{prop}

An $F$-coalgebra $(\Omega, \omega)$ is called \emph{final} whenever, for any $F$-coalgebra $(X, c)$, there is a unique homomorphism $\llbracket - \rrbracket$ from $(X, c)$ to $(\Omega, \omega)$. In general, this final homomorphism can be regarded as an interpretation function, interpreting elements of $F$-coalgebras as elements of $\Omega$.

When we are concerned with different $F$-coalgebras simultaneously, say $(X, c)$ and $(Y, d)$, and their respective mappings to the final coalgebra, we sometimes denote the respective homomorphisms with the name of the carrier of the coalgebra, e.g.~$\llbracket - \rrbracket_X$ and $\llbracket - \rrbracket_Y$.

As an example of a final coalgebra, consider the set $\mathcal{P}(A^*)$ of all languages on the alphabet $A$. This set can be equipped with a $\mathbb{B} \times (-)^A$-coalgebra mapping (or, equivalently: can be assigned the structure of an---infinite---deterministic automaton) by setting
\[
o(L) = \textbf{if $\epsilon \in L$ then $1$ else $0$}
\quad\textrm{and}\quad
L_a = \{ w \in A^* \,|\, aw \in L \}
\]
for every $L \in \mathcal{P}(A^*)$.

\begin{prop}\label{prop:finality}
The $\mathbb{B} \times (-)^A$-coalgebra mapping just defined on $\mathcal{P}(A^*)$ is final, and, given a $\mathbb{B} \times (-)^A$-coalgebra $(X, (o, \delta))$, the unique homomorphism $\llbracket - \rrbracket: X \to \mathcal{P}(A^*)$ is given by
\[
\llbracket x \rrbracket = \{ w \in A^* \,|\, o(x_w) = 1 \}
\]
for all $x \in X$.
\end{prop}

Similarly, for every set $K$, the set of all functions in $K^{A^*}$, or formal power series in noncommuting variables over $A$ and with coefficients in $K$, (often denoted by $K \llangle A \rrangle$) can be equipped with a $K \times (-)^A$-coalgebra map making it final ~\cite{Ru00}.

A \emph{bisimulation} between two $F$-coalgebras $(X, c)$ and $(Y, d)$ is a relation $R \subseteq X \times Y$ such that there exists a morphism $r: R \to FR$, making the diagram
\begin{diagram} X & \lTo_{\pi_1} & R & \rTo_{\pi_2} & Y \\ \dTo_c & & \dTo_r & & \dTo_d \\ FX & \lTo_{F\pi_1} & FR & \rTo_{F\pi_2} & FY
\end{diagram} commute. Here $\pi_1$ and $\pi_2$ are the projections from $R$ to $X$ and $Y$, respectively. Whenever there exists a bisimulation $R$ such that $(x, y) \in R$, we say that $x$ and $y$ are \emph{bisimilar} and write $x \sim y$. The relation $\sim$ itself again is a bisimulation, as well as an equivalence relation.

When $F$ preserves weak pullbacks (which is true in the case of functors of type $K \times (-)^A$) and a final $F$-coalgebra exists, we have
\[
x \sim y \qquad \textrm{iff} \qquad \llbracket x \rrbracket_X = \llbracket y \rrbracket_Y
\]
for all $x \in X$ and $y \in Y$. In other words, $x$ and $y$ are bisimilar exactly when they are mapped onto the element in the final coalgebra \cite{Ru00}. Again, these elements are formal languages in the case of the functor $\mathbb{B} \times (-)^A$, and formal power series in the more general case of functors of the type $K \times (-)^A$.

In the case of $K \times (-)^A$-coalgebras, the above categorical definition of a bisimulation corresponds to the following condition:
\begin{prop}
Given $K \times (-)^A$-coalgebras $(X, (o, \delta))$ and $(Y, (o', \delta'))$, a relation $R \subseteq X \times Y$ is a bisimulation iff, whenever $x \; R \; y$, we have $o(x) = o(y)$, and $x_a \; R \;  y_a$ for all $a \in A$.
\end{prop}

Often the notion of a bisimulation \emph{up to} some property or equivalence class comes in handy. For instance, given $K \times (-)^A$-coalgebras $(X, (o, \delta))$ and $(Y, (o', \delta'))$, we call a relation $R \subseteq X \times Y$ a \emph{bisimulation up to bisimilarity} if, whenever $x \; R \; y$, we have $o(x) = o(y)$, and for all $a \in A$, there are $x' \in X$ and $y' \in Y$ such that $x_a \sim x'$, $x' \; R \; y'$, and $y' \sim y_a$. Clearly $\sim$ is a bisimulation up to bisimilarity. Conversely:

\begin{prop}
For every bisimulation up to bisimilarity $R$ between $K \times (-)^A$-coalgebras $(X, (o, \delta))$ and $(Y, (o', \delta'))$, if $x \; R \; y$, then $x \sim y$.
\end{prop}

\proof
We extend $R$ to a relation $\hat{R}$ defined by:
\[
\hat{R} \colonequals \{ (x, y) \,|\, \exists x', y': x \sim x' \land x' \; R \; y' \land y' \sim y \}
\]
We verify that $\hat{R}$ is a bisimulation: first, if $x \; \hat{R} \; y$, then there are $x'$ and $y'$ such that $x \sim x'$, $x' \; R \; y'$, and $y' \sim y$, and hence $o(x) = o(x') = o(y') = o(y)$. From $x' \; R \; y'$, it follows that there are $x'' \in X$ and $y'' \in Y$ such that $x'_a \sim x''$, $x'' \; R \; y''$, and $y'' \sim y'_a$. However, from $x \sim x'$ and $y \sim y'$, respectively, we also obtain $x_a \sim x'_a$ and $y_a \sim y'_a$. By transitivity of $\sim$, we now obtain $x_a \sim x''$ and $y'' \sim y_a$, and hence it follows that $x_a \; \hat{R} \; y_a$, so $\hat{R}$ is a bisimulation.
\qed

For a more systematic account of bisimulations up to, we refer to \cite{RoBoRu12}.

\subsection{Context-free languages and grammars}
\label{sec:cfg}

We assume the reader to be familiar with the standard definitions pertaining to context-free grammars and languages, and give only the definitions and results we need in the rest of this article. For a more comprehensive treatment of context-free grammars and languages, see e.g.~\cite{HoMoUl06} or \cite{AuBeBo97}.

We will represent \emph{context-free grammar} on a finite alphabet $A$ as a pair $(X,p)$, where $X$ is a finite set of symbols we call \emph{nonterminals} or \emph{variables}, and
\[
p : X \to \mathcal{P}_\omega((A + X)^*)
\]
is a function describing the production rules. Here $+$ denotes the coproduct (or disjoint union), $\mathcal{P}_\omega$ the finite power set, and $(A+X)^*$ is the set of all words over the disjoint union $A + X$. We use the notation
\[
x \to w
\]
to denote $w \in p(x)$, where $x \in X$ and $w \in (A + X)^*$. Furthermore, the notation
\[
x \to w_1 \;|\; \ldots \;|\; w_n
\]
is used to denote $\{ w_1, \ldots, w_n \} \subseteq p(x)$.

Note that the finiteness conditions on both $X$ and the powerset are required, because otherwise the set of resulting languages would be the set of \emph{all} languages.

In order to define the language associated to a context-free grammar, we now introduce the notion of a derivation. Given a context-free grammar $(X,p)$, for all words $v, w \in (A + X)^*$, we write $v \Rightarrow w$, and say $w$ is derivable from $v$ in a single \emph{derivation} step, whenever $v = v_1xv_2$ and $w = v_1uv_2$ for a production rule $x \to u$, and $v_1,v_2 \in (A + X)^*$. We say that $w$ is derivable from $v$ in a single \emph{leftmost derivation} step whenever $v_1$ is a (possibly empty) word in $A^*$. As usual, $\Rightarrow^*$ denotes the reflexive and transitive closure of $\Rightarrow$. In general, if $v \Rightarrow^* w$, then $w$ is derivable from $v$ using only leftmost derivation steps. Therefore we can restrict our attention to leftmost derivations only.

For a context-free grammar $(X,p)$ and any variable $x \in X$, called the starting symbol, we define the language $\mathcal{L}(x) \in \mathcal{P}(A^*)$ generated by $(X,p)$ from $x$ as:
\[
\mathcal{L}(x) \colonequals \{ w \in A^* \,|\, x \Rightarrow^* w \}
\]

A language $L \in \mathcal{P}(A^*)$ is called \emph{context-free} if there exists a context free grammar $(X,p)$ and a variable $x \in X$, such that $L = \mathcal{L}(x)$.

For our coalgebraic treatment of context-free languages it will be convenient to work with context-free grammars with production rules of a specific form. We say that a context-free grammar is in \emph{Greibach normal form} if all of its production rules are of the form
\[
x \to aw \quad \textrm{or} \quad x \to \epsilon
\]
where $a \in A$ is an alphabet symbol, and $w \in X^*$ is a (possibly empty) sequence of nonterminal symbols.\footnote{We remark that, in \cite{WiBoRu11}, we used a variant called the weak Greibach normal form, differing from the familiar Greibach normal form~\cite{Gr65} in that the right hand side of production rules there can consist of words $w$ over $A + X$, rather than simply over $X$.} It is well-known (see e.g.~\cite{Gr65}) that for every context-free language $L$, there exists a context-free grammar $(X, p)$ in Greibach normal form, and some $x \in X$, such that $\mathcal{L}(x) = L$.

\section{Context-Free Languages via Grammar Coalgebras}
\label{sec:cfgcoalgebra}

In this section, we will represent context-free grammars in Greibach normal form (for a fixed finite alphabet $A$) using coalgebras for the functor $\mathbb{B} \times \mathcal{P}_\omega((-)^*)^A$, yielding a coalgebraic semantics for these grammars (and hence, for context-free languages in general). We do this by showing that the unique coalgebra homomorphism from the coalgebraic representation of the grammar to the final $\mathbb{B} \times (-)^A$-coalgebra of all languages maps nonterminal symbols precisely to the context-free languages they generate.

The key observation here is that every context-free grammar $(X,p)$ with productions in Greibach normal form can be seen as a coalgebra for the functor $\mathbb{B} \times \mathcal{P}_\omega((-)^*)^A$. More precisely, we represent the production rules by a map $(o, \delta) :X \to \mathbb{B} \times \mathcal{P}_\omega(X^*)^A$ defined for all nonterminal symbols $x \in X$ by
\[ 
o(x) = \textbf{if $x \to \epsilon$ then $1$ else $0$} \quad \textrm{and} \quad x_a = \{ w \in X^* \,|\, x \to a w \}
\]
writing as before $x_a$ for $\delta(x)(a)$. We call this coalgebra the \emph{grammar coalgebra} corresponding to the grammar $(X, p)$.

As a first example, consider the following grammar in Greibach normal form, over the alphabet $A = \{ a,b \}$ with nonterminal symbols $X = \{ x, y, z \}$ and productions:
\begin{eqnarray*}
x & \to & \epsilon \,|\, axz \,|\, byz \\
y & \to & \epsilon \,|\, byz \\
z & \to & a
\end{eqnarray*}
The language generated from $x$ is $\mathcal{L}(x) = \{ a^nb^ma^{m + n} \,|\, m, n \in \mathbb{N} \}$, while the language generated from $y$ is $\mathcal{L}(y) = \{ b^na^n \,|\, n \in \mathbb{N} \}$. In coalgebraic form, the above productions read as follows:
\begin{center}
\begin{tabular}{c|c|c}
output value & $a$-derivative & $b$-derivative \\
\hline
$o(x) = 1$ & $x_a = \{ xz \}$ & $x_b = \{ yz \}$ \\
$o(y) = 1$ & $y_a = \emptyset$ & $y_b = \{ yz \}$ \\
$o(z) = 0$ & $z_a = \{ \epsilon \}$ & $z_b = \emptyset$
\end{tabular}
\end{center}
We can regard systems of this type as \emph{systems of behavioural differential equations}, where each derivative is given as a set of words over $X$, or equivalently, as an element of $\mathcal{P}_\omega(X^*)$.

The coalgebra associated to each context-free grammar in Greibach normal form is not a proper deterministic automaton (i.e. a $\mathbb{B} \times (-)^A$-coalgebra) but rather a $\mathbb{B} \times \mathcal{P}_\omega((-)^*)^A$-coalgebra. However, we can turn it into a deterministic automaton by embedding the nonterminal symbols $X$ into $\mathcal{P}_\omega(X^*)$ using the assignment $\eta_X: X \to \mathcal{P}_\omega(X^*)$ mapping each $x \in X$ into the singleton set $\{ x \}$ (in which $x$ is seen as a word). In fact, we extend in a canonical manner each grammar coalgebra $(o, \delta): X \to \mathbb{B} \times \mathcal{P}_\omega(X^*)^A$ to a $\mathbb{B} \times (-)^A$-coalgebra $(\hat{o}, \hat{\delta}): \mathcal{P}_\omega(X^*) \to \mathbb{B} \times \mathcal{P}_\omega(X^*)^A$, which we will call the \emph{grammar automaton} generated from $(X, p)$, as follows: for each $S \in \mathcal{P}_\omega(X^*)$ we define the output value and derivatives inductively by
\begin{center}
\begin{tabular}{c|c|c}
$S$ & $\hat{o}(S)$ & $S_a\, (a \in A)$ \\
\hline $\{ \epsilon \}$ & $1$ & $\emptyset$ \\
\specialcell{ $\{ x w \}$ \\ $(x \in X, w \in X^*)$} & $o(x) \land \hat{o}(\{ w \})$ & $x_a  \{ w \} \cup i(o(x)) \{ w \}_a$ \\
\specialcell{$\bigcup_{i \le n} \{ w_i \}$ \\ $(n \in \mathbb{N}, w_i \in X^*)$} & $\bigvee_{i \le n} \hat{o}(\{ w_i \})$ & $\bigcup_{i \le n} \{ w_i \}_a$
\end{tabular}
\end{center}
Here $i: \mathbb{B} \to \mathcal{P}(X^*)$ is a mapping defined by $i(1) = \{ \epsilon \}$ and $i(0) = \emptyset$. It is clear that $i$ is a semiring morphism, and it follows directly from the semiring axioms that
\[
i(o) \cdot L = (\textbf{if $o = 1$ then $L$ else $\emptyset$}).
\]
We can now combine the $\mathbb{B} \times \mathcal{P}_\omega((-)^*)^A$-coalgebra $(X, (o, \delta))$ with the $\mathbb{B} \times (-)^A$-coalgebra $(\mathcal{P}_\omega(X^*), (\hat{o}, \hat{\delta}))$ and the unique homomorphism into the final coalgebra as in the following commuting diagram: 
\begin{diagram}
X & \rTo_{\{ - \}} & \mathcal{P}_\omega(X^*) & \rDotsto_{\llbracket - \rrbracket} & \mathcal{P}(A^*) \\
\dTo_{(o, \delta)} & \ldTo_{(\hat{o}, \hat{\delta})} & & & \dTo \\
\mathbb{B} \times \mathcal{P}_\omega(X^*)^A & & \rDotsto & & \mathbb{B} \times \mathcal{P}(A^*)^A
\end{diagram}
We remark that the leftmost triangle in this diagram commutes: using the definition scheme for $\hat{o}$ and $\hat{\delta}$, it is easily verified that for all $x \in X$, $o(x) = \hat{o}(\{ x \})$ and $\delta(x) = \hat{\delta}(\{ x \})$.

The inductive definition consisting of $\hat{o}(\{ xw \}) = o(x) \land \hat{o}(\{ w \})$ and $\{ xw \}_a = x_a \{ w \} \cup i(o(x)) \{ w \}_a$ can be seen as an instance of a \emph{product rule}. The following proposition establishes a more general version of the product rule, in which the role of the nonterminal $x$ is taken over by arbitrary words $s \in X^*$:

\begin{prop}\label{prop:prodrule1}
For all words $s, t \in X^*$, the equations
\[
\hat{o}(\{ st \}) = \hat{o}(\{ s \}) \land \hat{o}(\{ t \}) \qquad \textrm{and} \qquad \{ st \}_a = \{ s \}_a\{ t \} \cup i(\hat{o}(\{ s \}))\{ t \}_a
\]
hold in any system defined as above.
\end{prop}

\proof
Induction on the length of $s$.

If $|s| = 0$, then $s = \epsilon$ giving
\[
\hat{o}(\{ st \}) = \hat{o}(\{ t \}) = 1 \land \hat{o}(t) = \hat{o}(\{ s \}) \land \hat{o}(\{ t \})
\]
and
\[
\{ st \}_a = \{ \epsilon t \}_a = \{ t \}_a = \emptyset \{ t \} \cup \{ \epsilon \}\{ t \}_a = \{ s \}_a\{ t \} \cup i(\hat{o}(\{ s \}))\{ t \}_a.
\]

If $|s| > 0$, then $s = xu$ for some $x \in X$ and $t \in X^*$, and use the inductive hypothesis that
\[
\hat{o}(\{ ut \}) = \hat{o}(\{ u \}) \land \hat{o}(\{ t \}) \qquad \textrm{and} \qquad \{ ut \}_a = \{ u \}_a\{ t \} \cup i(\hat{o}(\{ u \}))\{ t \}_a
\]
and now observe
\begin{eqnarray*}
\hat{o}(\{ st \}) & = & \hat{o}(\{ xut \}) \\
& = & o(x) \land \hat{o}(\{ ut \}) \\
& = & o(x) \land \hat{o}(\{ u \}) \land \hat{o}(\{ t \}) \\
& = & \hat{o}(\{ xu \}) \land \hat{o}(\{ t \}) \\
& = & \hat{o}(\{ s \}) \land \hat{o}(\{ t \})
\end{eqnarray*}
and
\begin{eqnarray*}
\{ st \}_a & = & \{ xut \}_a \\
& = & \{ x_a \}\{ ut \} \cup i(o(x))\{ ut \}_a \\ 
& = & \{ x_a \}\{ ut \} \cup i(o(x))(\{ u \}_a\{ t \} \cup i(\hat{o}(\{ u \}))\{ t \}_a) \\ 
& = & \{ x_a \}\{ ut \} \cup i(o(x))\{ u \}_a\{ t \} \cup i(o(x))i(\hat{o}(\{ u \}))\{ t \}_a \\ 
& = & \{ xu \}_a\{ t \} \cup i(\hat{o}(\{ xu \}))\{ t \}_a \\
& = & \{ s \}_a\{ t \} \cup i(\hat{o}(\{ s \})\{ t \}_a,
\end{eqnarray*}
completing the proof.
\qed

The following version of the product rule is yet more general:

\begin{prop}\label{prop:prodrule2}
For all finite languages $S, T \in \mathcal{P}_\omega(X^*)$, the equations
\[
\hat{o}(ST) = \hat{o}(S) \land \hat{o}(T) \qquad \textrm{and} \qquad (ST)_a = S_aT \cup i(\hat{o}(S))T_a
\]
hold in any system defined as above.
\end{prop}

\proof
We have
\[
\hat{o}(ST) = \hat{o}(\{ st \,|\, s \in S \land t \in T \}) = \bigvee_{s \in S} \bigvee_{t \in T} \hat{o}(\{ st \}) = \bigvee_{s \in S} \hat{o}(\{ s \}) \land \bigvee_{t \in T} \hat{o}(\{ t \}) = \hat{o}(S) \land \hat{o}(T)
\]
and
\begin{eqnarray*}
(ST)_a & = & \{ st \,|\, s \in S \land t \in T \}_a \\
& = & \bigcup_{s \in S} \bigcup_{t \in T} \{ st \}_a \\
& = & \bigcup_{s \in S} \bigcup_{t \in T} \left( \{ s \}_a \{ t \} \cup i(\hat{o}(\{ s \})) \{ t \}_a \right) \\
& = & \bigcup_{s \in S} \bigcup_{t \in T} \{ s \}_a \{ t \} \cup \bigcup_{s \in S} \bigcup_{t \in T} i(\hat{o}(\{ s \})) \{ t \}_a \\
& = & S_aT \cup i(\hat{o}(S))T_a.\rlap{\hbox to 196 pt{\hfill\qEd}}
\end{eqnarray*}

\begin{lem}\label{lem:second}
Given a $\mathbb{B} \times \mathcal{P}_\omega((-)^*)^A$-coalgebra $(X, (o, \delta))$, $S \in \mathcal{P}_\omega(X^*)$, and $w \in A^*$, $\hat{o}(S_w) = \bigvee_{s \in S} \hat{o}(\{ s \}_w)$ w.r.t.~the extension $(\mathcal{P}_\omega(X^*), (\hat{o}, \hat{\delta}))$.
\end{lem}

\proof
Induction on the length of $w$.

If $|w| = 0$, then $w = \epsilon$, and hence
\[
\hat{o}(S_w) = \hat{o}(S) = \bigvee_{s \in S} \hat{o}(\{ s \}) = \bigvee_{s \in S} \hat{o}(\{ s \}_w)
\]
by definition.

If $|w| = k + 1$, then $w = av$ for some $a \in A$ and $v \in A^*$, and assume as inductive hypothesis that the stated property holds for all $v$ with $|v| \le k$. We now obtain
\[
\hat{o}(S_w) = \hat{o}((S_a)_v) = \bigvee_{t \in S_a} \hat{o}(\{ t \}_v) = \bigvee_{s \in S} \bigvee_{t \in \{ s \}_a} \hat{o}(\{ t \}_v) = \bigvee_{s \in S} \hat{o}((\{ s \}_a)_v) = \bigvee_{s \in S} \hat{o}(\{ s \}_w)
\]
and the proof is complete.
\qed

Given a context-free grammar $(X, p)$ in Greibach normal form, let $X_\mathrm{nul}$ denote the set
\[
\{ x \in X \,|\, x \to \epsilon \}
\]
of nullable nonterminals in $X$.

\begin{lem}\label{lem:nullable}
Given a $\mathbb{B} \times \mathcal{P}_\omega((-)^*)^A$-coalgebra $(X, (o, \delta))$, we have $\{ t \}_a \subseteq \{ st \}_a$ for $s \in (X_\mathrm{nul})^*$ and $t \in X^*$ w.r.t.~the extension $(\mathcal{P}_\omega(X^*), (\hat{o}, \hat{\delta}))$.
\end{lem}

\proof
From $s \in X^*_\mathrm{nul}$, we get $\hat{o}(\{ s \}) = 1$. We now get
\[
\{ t \}_a \subseteq \{ s \}_a \{ t \} \cup \{ t \}_a = \{ s \}_a \{ t \} \cup i(\hat{o}(\{ s \})) \{ t \}_a = \{ st \}_a
\]
using Proposition~\ref{prop:prodrule1}.
\qed

\begin{lem}\label{lem:derivation}
Let $(X, p)$ be a context-free grammar in Greibach normal form over a finite alphabet $A$. Let $s, t \in X^*$ and let $a \in A$. If $t \in \{ s \}_a$ (w.r.t.~the grammar automaton generated from $(X, p)$) , then also $s \Rightarrow^* at$.
\end{lem}

\proof
Assume that $t \in \{ s \}_a$. We proceed by induction on the length of $s$. If $|s| = 0$, we would obtain $t \in \{ \epsilon \}_a = \emptyset$, which cannot be the case, making the base case trivially true.

If $|s| > 0$, then $s = xu$ for some $x \in X$ and $u \in X^*$. From $t \in \{ s \}_a$ we now obtain
\[
t \in \{ vu \,|\, v \in x_a \} \cup i(o(x)) \{ u \}_a.
\]
Equivalently, either $t$ is of the form $vu$ for some $v \in x_a$, or $o(x) = 1$ and $t \in \{ u \}_a$. In the first case, from $v \in x_a$ we get $x \to av$ directly as a result of the definition, and hence $s = xu \Rightarrow avu = at$ and also $s \Rightarrow^* at$. In the second case, apply the inductive hypothesis to obtain $u \Rightarrow^* at$ from $t \in \{ u \}_a$, combine this with $x \to \epsilon$ (which we know because $o(x) = 1$), giving $s = xu \Rightarrow u \Rightarrow^* at$ and completing the proof.
\qed

The following proposition now establishes the correspondence between derivations in a context-free grammar in Greibach normal form and the behaviour of a grammar automaton:

\begin{prop}\label{prop:derivderiv}
Let $(X, p)$ be a context-free grammar in Greibach normal form over a finite alphabet $A$. Let $s \in X^*$ and let $w \in A^*$. We have $s \Rightarrow^* w$ iff $\hat{o}(\{ s \}_w) = 1$ w.r.t.~the grammar automaton generated from $(X, p)$.
\end{prop}

\proof
Induction on the length of $w$.

If $|w| = 0$, then $w = \epsilon$, and note that we have $s \Rightarrow^* \epsilon$ iff $s \in X_\mathrm{nul}^*$. Now, using induction on the length of $s$, we can show that $\hat{o}(\{ s \}) = 1$ if and only if $s \in X_\mathrm{nul}^*$, completing the base case as $\hat{o}(\{ s \}_\epsilon) = \hat{o}(\{ s \})$.

If $|w| > 0$ then $w = av$ for some $a \in A$ and $v \in A^*$, and now assume as inductive hypothesis that the statement of the proposition holds for all words of length less than $w$. For the left-to-right direction, take a leftmost derivation witnessing $s \Rightarrow^* av$. There must be a first stage in the derivation where a rule of the form $x \to au$ is applied. Because $(X, p$) is in Greibach normal form, this means that at earlier stages only rules of the form $y \to \epsilon$ (for $y \in X$) are used, allowing us to decompose the derivation as
\[
s = txz \Rightarrow^* xz \Rightarrow auz \Rightarrow^* av
\]
for some $t \in X^*_\mathrm{nul}$, $x \in X$, $u, z \in X^*$, $a \in A$ and $v \in A^*$.

Using the inductive hypothesis, we now get $\hat{o}(\{ uz \}_v) = 1$ from $uz \Rightarrow^* v$. By Lemma~\ref{lem:nullable}, we have $\{ xz \}_a \subseteq \{ s \}_a$. Observe
\[
\{ s \}_a \supseteq \{ xz \}_a = \{ vz \,|\, v \in x_a \} \cup i(o(x)) \{ z \}_a \supseteq \{ uz \}
\]
with the last inclusion holding because $u \in x_a$ as a result of $x \to au$. But now
\[
\hat{o}(\{ s \}_w) = \hat{o}((\{ s \}_a)_v) = \bigvee_{t \in \{ s \}_a} \hat{o}(t_v) > \hat{o}(\{ uz \}_v) = 1
\]
and hence also $\hat{o}(\{ s \}_w) = 1$.

For the right-to-left direction, assume $\hat{o}(\{ s \}_{av}) = 1$. We now get:
\[
\hat{o}(\{ s \}_{av}) = \hat{o}((\{ s \}_a)_v) = \hat{o}\left(\bigcup_{t \in \{ s \}_a} \{ t \}_v\right) = \bigvee_{t \in \{ s \}_a} \hat{o}(\{ t \}_v)
\]
Because $\bigvee_{t \in \{ s \}_a} \hat{o}(\{ t \}_v) = 1$, there must be a $t \in \{ s \}_a$ such that $\hat{o}(\{ t \}_v) = 1$. Applying the inductive hypothesis now gives $t \Rightarrow^* v$, and from $t \in \{ s \}_a$ we obtain $s \Rightarrow^* at$ by Lemma~\ref{lem:derivation}, giving
\[
s \Rightarrow^* at \Rightarrow^* av = w
\]
completing the proof.
\qed

\begin{prop}\label{prop:grammarcoalgebratheorem}
Let $(X,p)$ be a context-free grammar in Greibach normal form over a finite alphabet $A$, and
$S \in \mathcal{P}_\omega(X^*)$. We have
\[
\llbracket S \rrbracket = \bigcup_{s \in S} \{ w \in A^* \,|\, s \Rightarrow^* w \}
\]
w.r.t.~the grammar automaton generated from $(X, p)$.
\end{prop}

\proof
We have
\begin{eqnarray*}
\llbracket S \rrbracket & = & \{ w \,|\, \hat{o}(S_w) = 1 \} \qquad \textit{by Proposition~\ref{prop:finality}} \\
& = & \{ w \,|\, \bigvee_{s \in S} \hat{o}(\{s \}_w) = 1 \} \qquad \textit{by Lemma~\ref{lem:second}} \\
& = & \bigcup_{s \in S} \{ w \,|\, \hat{o}(\{ s \}_w) = 1 \} \\
& = & \bigcup_{s \in S} \{ w \,|\, s \Rightarrow^* w \} \qquad \textit{by Proposition~\ref{prop:derivderiv}}.\rlap{\hbox to 101 pt{\hfill\qEd}}
\end{eqnarray*}

\begin{thm}\label{thm:firstequivalence}
The following conditions are equivalent:
\begin{enumerate}[\em(1)]
\item A language $L$ is context-free.
\item There exists a $\mathbb{B} \times \mathcal{P}_\omega((-)^*)^A$-coalgebra $(X, (o, \delta))$ and an $x \in X$ such that $\llbracket \{ x \} \rrbracket = L$ w.r.t. the extension $(\mathcal{P}_\omega(X^*), (\hat{o}, \hat{\delta}))$.
\item There exists a $\mathbb{B} \times \mathcal{P}_\omega((-)^*)^A$-coalgebra $(X, (o, \delta))$ and a $S \in \mathcal{P}_\omega(X^*)$ such that $\llbracket S \rrbracket = L$  w.r.t. the extension $(\mathcal{P}_\omega(X^*), (\hat{o}, \hat{\delta}))$.
\end{enumerate}
\end{thm}

\proof
$(1) \Rightarrow (2):$ If $L$ is context-free, there exists a context-free grammar $(X, p)$ in Greibach normal form and a $x \in X$ such that $\mathcal{L}(x) = L$. We have $L = \{ w \in A^* \,|\, x \Rightarrow^* w \}$, and using Proposition~\ref{prop:grammarcoalgebratheorem}, we now obtain
\[
\llbracket \{ x \} \rrbracket = \{ w \in A^* \,|\, x \Rightarrow^* w \} = L
\]
w.r.t.~the grammar automaton generated from $(X, p)$.

$(2) \Rightarrow (3):$ Given $(X, (o, \delta))$, take $S = \{ x \}$, and we obtain $\llbracket S \rrbracket = \llbracket \{ x \} \rrbracket = L$.

$(3) \Rightarrow (1):$ Given $(X, (o, \delta))$, define a context-free grammar $(X, p)$ in Greibach normal form by setting
\[
p(x) = i(o(x)) \cup \left( \bigcup_{a \in A} \{ a \} x_a \right)
\]
for all $x \in X$. It is easily checked that the construction from this section, starting from $(X, p)$, yields back $(X, (o, \delta))$. Now, Proposition~\ref{prop:grammarcoalgebratheorem} gives $\llbracket S \rrbracket = \bigcup_{s \in S} \{ w \in A^* \,|\, s \Rightarrow^* w \}$. Construct a new grammar (generally not in Greibach normal form) $(X \cup \{ x_0 \}, p')$ (with $x_0$ not occurring in $X$) and set $p'(x) = p(x)$ for all $x \in X$ and $p(x_0) = S$. We now have
\[
\mathcal{L}(x_0) = \{ w \in A^* \,|\, x_0 \Rightarrow^* w \} = \bigcup_{s \in S} \{ w \in A^* \,|\, s \Rightarrow^* w \} = \llbracket S \rrbracket = L
\]
with the second equality holding because the first derivation step used  in a derivation $x_0 \Rightarrow^* w$ must be of the form $x_0 \to s$ for some $s \in S$. From $\mathcal{L}(x_0) = L$, it follows directly that $L$ is context-free.
\qed

\section{Context-Free Languages via Syntactic Systems of Behavioural Differential Equations}
\label{sec:langeq}

We will now take a look at another coalgebraic characterization of context-free languages, where derivatives are given as elements of a term algebra.

Assuming a fixed finite alphabet $A$, and given a finite set of nonterminals $X$, we let the set $\mathcal{T}(X)$ of terms over $X$ be defined by the following specification:
\[
\tau \coloncolonequals \bar{0} \,|\, \bar{1} \,|\, \bar{x} \; (x \in X) \,|\, \bar{a} \; (a \in A) \,|\, \tau + \tau \,|\, \tau \times \tau
\]
Examples of terms include $\bar{1}$, $\bar{a} + \bar{x}$, $\bar{x} + \bar{a}$, and $(\bar{a} \times \bar{x}) + \bar{b}$. We note that these terms are purely syntactic, and assume no precedence rules: in order to disambiguate, we always will use parentheses. In this section, we will moreover carefully distinguish variables $x$ and their corresponding terms $\bar{x}$.

Given a finite set $X$ of nonterminals, a \emph{syntactic system of behavioural differential equations} is a coalgebra of the type
\[
X \stackrel{(o, \delta)}{\to} \mathbb{B} \times \mathcal{T}(X)^A
\]
assigning to each nonterminal an output value, and a function from alphabet symbols $a$ to $a$-derivatives, given as terms. Following the earlier convention, we again write $x_a$ for $\delta(x)(a)$.

As an example of a syntactic system of behavioural equations, over the set $\{ x, y \}$ of nonterminals, consider the following system:
\begin{center}
\begin{tabular}{c|c|c}
output value & $a$-derivative & $b$-derivative \\
\hline \\ [-2.4ex]
$o(x) = 1$ & $x_a = \bar{x} \times \bar{a}$ & $x_b = \bar{y} \times \bar{a}$ \\
$o(y) = 1$ & $y_a = \bar{0}$ & $y_b = \bar{y} \times \bar{a}$ \\
\end{tabular}
\end{center}
We will later see that this system is equivalent to the $\mathbb{B} \times (\mathcal{P}_\omega(-)^*)^A$-coalgebra used as an example in the previous section, with $x$ being interpreted as the language $\{ a^nb^ma^{m + n} \,|\, m, n \in \mathbb{N} \}$, and $y$ being interpreted as the language $\{b^na^n \,|\, n \in \mathbb{N} \}$.

Again, we will extend systems of equations $(o, \delta): X \to \mathcal{T}(X)^A$ to $\mathbb{B} \times (-)^A$-coalgebras
\[
(\bar{o}, \bar{\delta}): \mathcal{T}(X) \to \mathbb{B} \times \mathcal{T}(X)^A,
\]
by inductively defining the value of the mappings $\bar{o}$ and $\bar{\delta}$ on all terms $\tau \in \mathcal{T}$ (and alphabet symbols $a \in A$) as follows:
\begin{center}
\begin{tabular}{c|c|c}
$\tau$ & $\bar{o}(\tau)$ & $\tau_a$ \\
\hline
$\bar{x} \; (x \in X)$ & $o(x)$ & $x_a$ \\
$\bar{0}$ & $0$ & $\bar{0}$ \\
$\bar{1}$ & $1$ & $\bar{0}$ \\
$\bar{b} \; (b \in A)$ & $0$ & $j((b = a)\textbf{?})$ \\
$\sigma + \upsilon$ & $\bar{o}(\sigma) \lor \bar{o}(\upsilon)$ & $\sigma_a + \upsilon_a$ \\
$\sigma \times \upsilon$ & $\bar{o}(\sigma) \land \bar{o}(\upsilon)$ & $(\sigma_a \times \upsilon) + (j(\bar{o}(\sigma)) \times \upsilon_a)$
\end{tabular}
\end{center}\vspace{2 pt}

\noindent Here, the mapping $j: \mathbb{B} \to \mathcal{T}(X)$ is defined by $j(0) = \bar{0}$ and $j(1) = \bar{1}$, and $(b = a)\textbf{?}$ is defined as $\textbf{if $b = a$ then $1$ else $0$}$.

We can, again, combine the coalgebras $(X, (o, \delta))$ and $(\mathcal{T}(X), (\bar{o}, \bar{\delta}))$ in the following commuting diagram, together with the final homomorphism from $(\mathcal{T}(X), (\bar{o}, \bar{\delta}))$ to the final coalgebra:

\begin{diagram}
X & \rTo_{\eta_X} & \mathcal{T} (X) & \rDotsto_{\llbracket - \rrbracket} & \mathcal{P}(A^*) \\
\dTo_{(o, \delta)} & \ldTo_{(\bar{o}, \bar{\delta})} & & & \dTo \\
\mathbb{B} \times \mathcal{T}(X)^A & & \rDotsto & & \mathbb{B} \times \mathcal{P}(A^*)^A
\end{diagram}\vspace{2 pt}

\noindent We again call the composition $\llbracket \eta_X(-) \rrbracket: X \to \mathcal{P}(A^*)$ of the final homomorphism $\llbracket - \rrbracket$ with the injection $\eta_X$ of the monad $\mathcal{T}$ the \emph{solution} to the system $(X, (o, \delta))$.

We will now, in order to transform $\mathbb{B} \times \mathcal{T}(-)^A$-coalgebras into $\mathbb{B} \times \mathcal{P}_\omega((-)^*)^A$-coalgebras, inductively define a mapping $f: \mathcal{T}(X) \to \mathcal{P}_\omega(Y^*)$, where
\[
Y \colonequals \{ \hat{x} \,|\, x \in X \} \cup \{ \hat{a} \,|\, a \in A \}
\]
consists of new notational variants of the nonterminals $x \in X$ as well as the alphabet symbols $a \in A$. The mapping $f$ is defined as follows:

\begin{center}
\begin{tabular}{c|c}
$\tau$ & $f(\tau)$ \\
\hline
$\bar{x} \; (x \in X)$ & $\{ \hat{x} \}$ \\
$\bar{0}$ & $\emptyset$ \\
$\bar{1}$ & $\{ \epsilon \}$ \\
$\bar{a} \; (a \in A)$ & $\{ \hat{a} \}$ \\
$\sigma + \upsilon$ & $f(\sigma) \cup f(\upsilon)$ \\
$\sigma \times \upsilon$ & $f(\sigma)f(\upsilon)$
\end{tabular}
\end{center}
Now, say, we are given a syntactic system of behavioural differential equations
\[
(o, \delta): X \to \mathbb{B} \times \mathcal{T}(X)^A.
\]
We now construct a $\mathbb{B} \times \mathcal{P}((-)^*)^A$-coalgebra
$(Y, (o', \delta'))$ by specifying:
\begin{center}
\begin{tabular}{c|c|c}
$y \in Y$ & $o'(y)$ & $y_a$ \\
\hline
$\hat{x} \; (x \in X)$ & $o(x)$ & $f(x_a)$ \\
$\hat{b} \; (b \in A)$ & $0$ & $i((b = a)\textbf{?})$ \\
\end{tabular}
\end{center}

The following lemma is entirely trivial but worth stating nonetheless:
\begin{lem}
For all $k \in \mathbb{B}$, $f(j(k)) = i(k)$.
\end{lem}

\proof
Check the definitions of $f$, $i$ and $j$.
\qed

\begin{prop}
The mapping $f$ is a $\mathbb{B} \times (-)^A$-coalgebra homomorphism or, in other words, the diagram
\begin{diagram}
\mathcal{T}(X) & \rTo_f & \mathcal{P}_\omega(Y^*) \\
\dTo_{(\bar{o}, \bar{\delta})} & & \dTo_{(\hat{o}, \hat{\delta})} \\
\mathbb{B} \times \mathcal{T}(X)^A & \rTo & \mathbb{B} \times \mathcal{P}_\omega(Y^*)^A
\end{diagram}
commutes.
\end{prop}

\proof
In order to show that $f$ is a homomorphism, we need to show that for all terms $\sigma \in \mathcal{T}(X)$,
\[
\hat{o}(f(\sigma)) = \bar{o}(\sigma) \qquad \textrm{and} \qquad f(\sigma)_a = f(\sigma_a).
\]
We do this by structural induction on terms $\sigma \in \mathcal{T}(X)$.

Base cases:
\begin{iteMize}{$\bullet$}
\item Case $\sigma = \bar{x} \; (x \in X)$: $\hat{o}(f(\bar{x})) = \hat{o}(\{ \hat{x} \}) = o'(\hat{x}) = o(x) = \bar{o}(\bar{x})$ and $f(\bar{x})_a = \{ \hat{x} \}_a = \hat{x}_a = f(x_a) = f(\bar{x}_a)$.
\item Case $\sigma = \bar{b} \; (b \in A)$: $\hat{o}(f(\bar{b})) = \hat{o}(\{ \hat{b} \}) = o'(\hat{b}) = 0 = \bar{o}(\bar{b})$ and $f(\bar{b})_a = \{ \hat{b} \}_a = \hat{b}_a = i((b = a)\textbf{?}) = f(j(b = a)\textbf{?}) = f(\bar{b}_a)$.
\item Case $\sigma = \bar{1}$: $\hat{o}(f(\bar{1})) = \hat{o}(\{ \epsilon \}) = 1 = \bar{o}(\bar{1})$ and $f(\bar{1})_a = \{ \epsilon \}_a = \emptyset = f(\bar{0}) = f(\bar{1}_a)$.
\item Case $\sigma = \bar{0}$: $\hat{o}(f(\bar{0})) = \hat{o}(\emptyset) = 0 = \bar{o}(\bar{0})$ and $f(\bar{0})_a = \emptyset_a = \emptyset = f(\bar{0}) = f(\bar{0}_a)$.
\end{iteMize}

For the inductive cases, use the inductive hypothesis that $\hat{o}(f(\tau)) = \bar{o}(\tau)$, $\hat{o}(f(\upsilon)) = \bar{o}(\upsilon)$, $f(\tau)_a = f(\tau_a)$, and $f(\upsilon)_a = f(\upsilon_a)$ to show that the homomorphic property is satisfied for terms $\tau + \upsilon$ and $\tau \times \upsilon$:

\begin{iteMize}{$\bullet$}
\item Case $\sigma = \tau + \upsilon$: we have
\[
\hat{o}(f(\tau + \upsilon)) = \hat{o}(f(\tau) \cup f(\upsilon)) = \hat{o}(f(\tau)) \lor \hat{o}(f(\upsilon)) = \bar{o}(\tau) \lor \bar{o}(\upsilon) = \bar{o}(\tau + \upsilon)
\]
and
\[
f(\tau + \upsilon)_a = (f(\tau) \cup f(\upsilon))_a = f(\tau)_a \cup f(\upsilon)_a = f(\tau_a) \cup f(\upsilon_a) = f((\tau + \upsilon)_a).
\]
\item Case $\sigma = \tau \times \upsilon$: we have
\[
\hat{o}(f(\tau \times \upsilon)) = \hat{o}(f(\tau)f(\upsilon)) = \hat{o}(f(\tau)) \land \hat{o}(f(\upsilon)) = \bar{o}(\tau) \land \bar{o}(\upsilon) = \bar{o}(\tau \times \upsilon)
\]
and
\begin{eqnarray*}
f(\tau \times \upsilon)_a & = & (f(\tau)f(\upsilon))_a \\
& = & f(\tau)_af(\upsilon) \cup i(\hat{o}(f(\tau))f(\upsilon)_a \\
& = & f(\tau_a)f(\upsilon) \cup i(\bar{o}(\tau))f(\upsilon_a) \\
& = & f(\tau_a)f(\upsilon) \cup f(j(\bar{o}(\tau)))f(\upsilon_a) \\
& = & f(\tau_a \times \upsilon) \cup f(j(\bar{o}(\tau)) \times \upsilon_a) \\
& = & f((\tau_a \times \upsilon) + (j(\bar{o}(\tau)) \times \upsilon_a)).
  \rlap{\hbox to 111 pt{\hfill\qEd}}
\end{eqnarray*}
\end{iteMize}

\noindent Moreover, $f$ is surjective and has a right-inverse $g$:
\begin{prop}
There is a mapping $g: \mathcal{P}_\omega(Y^*) \to \mathcal{T}(X)$ such that $f \circ g = 1_{\mathcal{P}_\omega(Y^*)}$. Hence, $f$ is surjective.
\end{prop}

We now are able to state the second equivalence theorem:
\begin{thm}\label{thm:secondequivalence}
The following conditions are equivalent:
\begin{enumerate}[\em(1)]
\item A language $L$ is context-free.
\item There exists a $\mathbb{B} \times \mathcal{T}(-)^A$-coalgebra $(X, (o, \delta))$ and an $x \in X$ such that $\llbracket \eta_X(x) \rrbracket = L$ w.r.t. the extension $(\mathcal{T}(X), (\bar{o}, \bar{\delta}))$.
\item There exists a $\mathbb{B} \times \mathcal{T}(-)^A$-coalgebra $(X, (o, \delta))$ and a $\tau \in \mathcal{T}(X)$ such that $\llbracket \tau \rrbracket = L$ w.r.t. the extension $(\mathcal{T}(X), (\bar{o}, \bar{\delta}))$.
\end{enumerate}
\end{thm}

\proof
$(1) \Rightarrow (2)$: Assume $L$ is context-free. By Theorem~\ref{thm:firstequivalence}, there exists a $\mathbb{B} \times \mathcal{P}((-)^*)^A$-coalgebra $(\hat{X}, (o^0, \delta^0))$ and a $\hat{x} \in \hat{X}$ such that $\llbracket \eta_{\hat{X}}(\hat{x}) \rrbracket = L$. (We assume the carrier of this coalgebra to be of the form $\hat{X} = \{ \hat{x} \,|\, x \in X \}$ for the sake of notational convenience.)

We first construct another $\mathbb{B} \times \mathcal{P}((-)^*)^A$-coalgebra $(Y, (o^1, \delta^1))$, with $Y = \hat{X} \cup \{ \hat{a} \,|\, a \in A \}$. Set $o^1(\hat{x}) = o^0(\hat{x})$, $o^1(\hat{a}) = 0$, $\delta^1(\hat{x})(a) = \delta^0(\hat{x})(a)$, and $\delta^1(\hat{b})(a) = i((b = a)\textbf{?})$. We obtain $\llbracket \eta_Y(\hat{x}) \rrbracket = L$ as the new system is equivalent to the old system, only with additional (unused) notational variants for the alphabet symbols.

Now construct a $\mathbb{B} \times \mathcal{T}(-)^A$-coalgebra $(X, (o^2, \delta^2))$, defined by $o^2(x) = o^1(\hat{x})$ and $\delta^2(x)(a) = g(\delta^1(\hat{x})(a))$. The construction from earlier in this section now gives yet another system $(Y, (o^3, \delta^3))$, defined by $o^3(\hat{x}) = o^2(x)$, $o^3(\hat{a}) = 0$, $\delta^3(\hat{x})(a) = f(\delta^2(x)(a))$, and $\delta^3(\hat{b})(a) = i((b = a)\textbf{?})$. From $o^2(x) = o^1(\hat{x})$, $o^1(\hat{a}) = 0$, $f(\delta^2(x)(a)) = f(g(\delta^1(\hat{x})(a))) = \delta^1(\hat{x})(a)$, and $\delta^1(\hat{b})(a) = i((b = a)\textbf{?})$, it follows that $o^1 = o^3$ and $\delta^1 = \delta^3$, and hence $\llbracket \eta_X(x) \rrbracket = \llbracket \bar{x} \rrbracket = \llbracket f(\bar{x}) \rrbracket = \llbracket \{ \hat{x} \} \rrbracket = \llbracket \eta_{Y}(\hat{x}) \rrbracket = L$.

$(2) \Rightarrow (3)$: Given $(X, (o, \delta))$, take $\tau = \bar{x}$, and we obtain $\llbracket \tau \rrbracket = \llbracket \eta_X(x) \rrbracket = L$.

$(3) \Rightarrow (1)$: Given $(X, (o, \delta))$, use the construction presented in this section to obtain a $\mathbb{B} \times \mathcal{P}_\omega((-)^*)^A$-coalgebra $(Y, (o', \delta'))$, yielding a coalgebra morphism $f$ from the extension $(\mathcal{T}(X), (\bar{o}, \bar{\delta}))$ to the extension $(\mathcal{P}_\omega(Y^*), (\hat{o}, \hat{\delta}))$. Because of the uniqueness of the morphism into the final coalgebra, we get $\llbracket - \rrbracket_{\mathcal{T}(X)} = \llbracket - \rrbracket_{\mathcal{P}_\omega(Y^*)} \circ f$. But by Theorem~\ref{thm:firstequivalence}, from $\llbracket f(\tau) \rrbracket_{\mathcal{P}_\omega(Y^*)} = L$ it follows that $L$ is context-free.
\qed

\section{Generalizing the Notion of Context-Freeness}
\label{sec:generalize}

It turns out that the approaches from the previous two sections, as well as the main constructions and almost all of the propositions and theorems, can without difficulty be generalized from outputs in the semiring $\mathbb{B}$ to outputs in any commutative\footnote{For noncommutative semirings, defining an appropriate semiring structure on $K \langle X \rangle$ is problematic.} semiring $K$.

In order to get the right grip on this generalization, first observe that the set $\mathcal{P}(A^*)$ of all languages can also, equivalently, be regarded as the function space $A^* \to \mathbb{B}$ assigning a value of either $0$ or $1$ to every word $w \in A^*$.

If we, instead, consider mappings
\[
A^* \to K
\]
assigning a value taken from the semiring $K$ to every word $w \in A^*$, we obtain $K$-weighted languages, or equivalently, formal power series over noncommuting variables with coefficients in $K$. This space can be represented e.g.~using the notation $K^{A^*}$; in classical texts on automata theory, it is usually represented as either $K \llangle A \rrangle$ or $K \llangle A^* \rrangle$; for the remainder of this article, we will stick to $K \llangle A \rrangle$.

We can now add a final $K \times (-)^A$-coalgebra structure on $K \llangle A \rrangle$ by setting, for any $\sigma \in K \llangle A \rrangle$:
\[
o(\sigma) = \sigma(\epsilon) \qquad \textrm{and} \qquad \sigma_a = \lambda w.(\sigma(aw)).
\]

In Section~\ref{sec:cfgcoalgebra}, derivatives were presented as elements of $\mathcal{P}_\omega(X^*)$, or, in other words, as finite languages over the alphabet of nonterminals. These can be regarded as mappings
\[
\sigma: X^* \to \mathbb{B}
\]
with finite support, i.e.~with only finitely many $w \in X^*$ such that $\sigma(w) \ne 0$. Again following classical notation, we let $K \langle X \rangle$ denote the set of functions $X^*$ to $K$ with finite support, or equivalently the set of (noncommuting) polynomials over $X$ with coefficients in $K$. Polynomials can be represented using familiar notation, with e.g.
\[
1 + 2xy + 3xyz \in \mathbb{N} \langle \{ x, y, z \} \rangle
\]
representing a function mapping $\epsilon$ to $1$, $xy$ to $2$, $xyz$ to $3$, and all other words to $0$, and form a semiring with this standard addition and multiplication. The polynomial $1$ now takes over the role of the language $\{ \epsilon \}$ as the unit of the semiring of polynomials. Note that $K \langle X \rangle$ generalizes $\mathcal{P}_\omega(X^*)$ in the same way as $K \llangle A \rrangle$ generalizes $\mathcal{P}(A^*)$.

We now transform $K \times K \langle - \rangle^A$-coalgebras $X \stackrel{o, \delta}{\to} K \times K \langle X \rangle^A$ into $K \times (-)^A$-coalgebras $K \langle X \rangle \stackrel{\hat{o}, \hat{\delta}}{\to} K \times K \langle X \rangle^A$ using the following inductive scheme:

\begin{center}
\begin{tabular}{c|c|c}
$\sigma$ & $\hat{o}(\sigma)$ & $\sigma_a\, (a \in A)$ \\
\hline $1$ & $1$ & $0$ \\
\specialcell{ $x w$ \\ $(x \in X, w \in X^*)$} & $o(x) \cdot \hat{o}(\{ w \})$ & $x_a w + i(o(x)) w_a$ \\
\specialcell{$\sum_{i \le n} k_i w_i$ \\ $(n \in \mathbb{N}, w_i \in X^*, k_i \in K)$} & $\sum_{i \le n} k_i \hat{o}(w_i)$ & $\sum_{i \le n} (w_i)_a$
\end{tabular}
\end{center}
Again, this construction can be summarized using the following commuting diagram:

\begin{diagram}
X & \rTo_{\eta} & K \langle X \rangle & \rDotsto_{\llbracket - \rrbracket} & K \llangle A \rrangle \\
\dTo_{(o, \delta)} & \ldTo_{(\hat{o}, \hat{\delta})} & & & \dTo \\
K \times K \langle X \rangle^A & & \rDotsto & & K \times K \llangle A \rrangle^A
\end{diagram}
The class of power series that can be characterized by systems of this type (for finite $X$) corresponds precisely to the \emph{constructively algebraic} power series \cite{BoRuWi12}, which are the usual generalization of the context-free languages. We remark, however, that this general equivalence is proven by another method than the one presented in Section~\ref{sec:cfgcoalgebra}.

As an example of a context-free power series (over the semiring of the natural numbers and over a singleton alphabet), taken from \cite{Ru02}, consider the stream defined by the following system, consisting of just a single equation
\[
o(x) = 1 \qquad x' = \, x \cdot x \label{eq:catalan}
\]
where $x'$ denotes the derivative with respect to the single alphabet symbol, and $\cdot$ denotes the multiplication in the semiring of polynomials.

Its solution turns out to be the stream of Catalan numbers
\[
1, 1, 2, 5, 14, 42, 132, 429, 1430, \ldots
\]
of which $n$th element is a count of the number of well-bracketed words consisting of $n$ pairs of opening and closing brackets. For more background on these types of streams, concrete constructions as well as the coalgebraic view on the connection to counting functions of unambiguous grammars, we refer to \cite{BoRuWi12}.

The approach from Section~\ref{sec:langeq} can straightforwardly be generalized to commutative semirings. The language of terms $\mathcal{T}_K(X)$ now becomes
\[
\tau \coloncolonequals \bar{k} \; (k \in K) \,|\, \bar{x} \; (x \in X) \,|\, \bar{a} \; (a \in A) \,|\, \tau + \tau \,|\, \tau \times \tau
\]
and the inductive specification of derivatives of terms now becomes
\begin{center}
\begin{tabular}{c|c|c}
$\tau$ & $\bar{o}(\tau)$ & $\tau_a$ \\
\hline
$\bar{x} \; (x \in X)$ & $o(x)$ & $x_a$ \\
$\bar{k} \; (k \in K)$ & $s$ & $\bar{0}$ \\
$\bar{b} \; (b \in A)$ & $0$ & $j((b = a)\textbf{?})$ \\
$\sigma + \upsilon$ & $\bar{o}(\sigma) + \bar{o}(\upsilon)$ & $\sigma_a + \upsilon_a$ \\
$\sigma \times \upsilon$ & $\bar{o}(\sigma) \cdot \bar{o}(\upsilon)$ & $(\sigma_a \times \upsilon) + (j(\bar{o}(\sigma)) \times \upsilon_a)$
\end{tabular}
\end{center}
yielding the following commuting diagram:
\begin{diagram}
X & \rTo_{\eta} & \mathcal{T}_K(X) & \rDotsto_{\llbracket - \rrbracket} & K \llangle A \rrangle \\
\dTo_{(o, \delta)} & \ldTo_{(\bar{o}, \bar{\delta})} & & & \dTo \\
K \times \mathcal{T}_K(X)^A & & \rDotsto & & K \times K \llangle A \rrangle^A
\end{diagram}\vspace{6 pt}

\noindent We remark that all the results from Section~\ref{sec:langeq} can be generalized, with minimal modifications, to commutative semirings.

For a more in-depth perspective on these context-free power series, and their relation to automatic sequences and (weighted) languages, see \cite{BoRuWi12}.

\section{Context-Free Languages and Power Series via $\mu$-expressions}
\label{sec:cfe}

In this section, we will introduce guarded $\mu$-expressions as an extension of regular expressions, where the Kleene star is replaced by a (unique) fixed point operator $\mu$. We do this directly in the generalized form for commutative semirings. We will then specify a $K \times (-)^A$-coalgebra structure on the class of these expressions, and prove that the languages (respectively, power series) characterizable by such expressions are precisely the context-free languages (respectively, constructively algebraic series). In contrast to the coalgebraic approaches from the earlier sections, this formalism gives us a \emph{single} coalgebra of which the elements are mapped exactly to the context-free languages by the final homomorphism.

Our usage of fixed point expressions with a coinductive semantics has a very similar flavour to that in \cite{Si10}, in which fixed point expressions are used as a characterization of regular expressions over a variety of functors. The additional expressive power obtained by the context-free expressions presented here is due to an explicit inclusion of a concatenation operator.\footnote{In \cite{Si10}, a translation from the familiar format of regular expressions (with concatenation) into $\mu$-style expressions is given by means of substitution. However, this translation does not work for expressions of the type $x \cdot t$.} This provides an additional perspective on the treatment given here, in which `context-freeness' is obtained by the addition of a new operator to a calculus of regular expressions\footnote{Although this calculus does not explicitly contain the Kleene star, it can easily be expressed up to behavioural equivalence by means of the equality $t^* = \mu x . (1 + (\sum_{a \in A}(\bar{a} \times (t_a \times x))))$}, and may pave the way for an investigation of (1) extending this approach to other coinductively defined operators, and (2) extending this approach to a generalized notion of context-freeness for other functors.

We define the set of terms $t$ (henceforth to be called \emph{$\mu$-expressions}) and guarded terms $g$ over a commutative semiring $K$, an alphabet $A$ and a set of variables $X$ as follows\footnote{Here, we can assume $X$ to be any countably infinite set.}:
\begin{eqnarray*}
t & \coloncolonequals & \bar{k} \; (k \in K) \,|\, x \in X \,|\, \bar{a} \; (a \in A) \,|\, t + t \,|\, t \times t \,|\, \mu x . g \\
g & \coloncolonequals & \bar{a} \times t \; (a \in A) \,|\, \bar{k} \; (k \in K) \,|\, g + g
\end{eqnarray*}
We now let $\mathcal{T}_\mu$ denote the set of all closed $\mu$-expressions, and $\mathcal{T}^-_\mu$ the set of all $\mu$-expressions. Similarly, we let $\mathcal{T}_\gamma$ denote the set of all closed and guarded $\mu$-expressions, and $\mathcal{T}^-_\gamma$ the set of all guarded $\mu$-expressions.

We can now assign a $\mathbb{B} \times (-)^A$-coalgebra structure on the set $\mathcal{T}_\mu$ of closed $\mu$-expressions, by presenting a mapping assigning to each $t \in \mathcal{T}_\mu$ an output value $o(t) \in \mathbb{B}$, and the derivative $t_a \in \mathcal{T}_\mu$ for each alphabet symbol $a$. We can do this by extending the earlier inductive scheme defining the output values and derivatives of expressions with the new $\mu$-operator, which performs the role of an unfolding operator here:
\begin{center}
\begin{tabular}{c|c|c}
$t$ & $o(t)$ & $t_a$ \\
\hline
\rule{0pt}{2.2ex}$\bar{k} \; (k \in K)$ & $k$ & $\bar{0}$ \\
$\bar{b} \; (b \in A)$ & $0$ & $j((b = a)\textbf{?})$ \\
$u + v$ & $o(u) + o(v)$ & $u_a+v_a$\\
$u \times v$ & $o(u) \cdot o(v)$ & $(u_a \times v) + (j(o(u)) \times v_a)$ \\
$\mu x . u$ & $o(u[\mu x . u / x])$ & $(u[\mu x . u / x])_a$
\end{tabular}
\end{center}
Here $t[\mu x.u/x]$, as usual, denotes the term obtained from $t$ by replacing all free occurrences of $x$ by $\mu x. u$. As usual in the case of uniform substitution, this process normally also would include a renaming of free variables occurring in the term $\mu x.u$: however, because $\mu x.u$ is a closed expression, no renaming is needed. Because of the guardedness conditions of terms occurring directly inside the $\mu$ operator, it is easy to see that the above scheme indeed defines a mapping $(o, \delta): \mathcal{T}_\mu \to K \times (\mathcal{T}_\mu)^A$.

We will next, using purely coalgebraic techniques, show that the quotient $\mathcal{T}_\mu / \sim$ forms an idempotent semiring. In order to be able to do this, however, we first need the notion of bisimulation up to addition.

We say a relation $R \subseteq \mathcal{T}_\mu \times \mathcal{T}_\mu$ is a \emph{bisimulation up to $+$} iff, whenever $(r_1, r_2) \in R$, we have $o(r_1) = o(r_2)$ and furthermore, for all $a \in A$, there are $s, t, u, v$ such that $(r_1)_a = s + u$, $(r_2)_a = t + v$, $s \; R \; t$ and $u \; R \; v$. We will let the latter property be denoted by $s + u \; R + R \; t + v$. The following proposition establishes the soundness of bisimulation up to $+$:

\begin{prop}
If $R$ is a bisimulation up to $+$, then $R \subseteq \sim$.
\end{prop}

\proof
We extend $R$ to a new relation $\hat{R}$, defined inductively by
\[
\hat{R}_0 \colonequals R
\quad
\textrm{and}
\quad
\hat{R}_{k + 1} \colonequals \{ (s + u, t + v) \,|\, s \; \hat{R}_k \; t \land u \; \hat{R}_k v \} = \hat{R}_k + \hat{R}_k
\]
where $\hat{R} = \bigcup_{n \in \mathbb{N}} \hat{R}_n$.

We now will show, using induction on $k$, that if $(r_1, r_2) \in \hat{R}_k$, then $o(r_1) = o(r_2)$ and $((r_1)_a, (r_2)_a) \in \hat{R}_{k + 1}$. The base case directly follows from the fact that $R$ is a bisimulation up to $+$. Assume the property holds for all $m \le k$, and assume $(r_1, r_2) \in \hat{R}_{k + 1}$. Then there are $s, u, t, v$ with $r_1 = s + u$ and $r_2 = t + v$ such that $(s, t) \in \hat{R}_k$ and $(u, v) \in \hat{R}_k$. The inductive hypothesis now gives $(s_a, t_a) \in \hat{R}_{k + 1}$ and $(u_a, v_a) \in \hat{R}_{k + 1}$. We now get $o(r_1) = o(s) + o(u) = o(t) + o(v) = o(r_2)$, and
\[
(r_1)_a = s_a + u_a \; \hat{R}_{k + 2} \; t_a + v_a = (r_2)_a.
\]
By the definition of $\hat{R}$, it now directly follows that $\hat{R}$ is a bisimulation. Because $R \subseteq \hat{R}$, we now also get $R \subseteq \sim$.
\qed

Again, the bisimilarity relation $\sim$ is a congruence with respect to the sum $+$ and multiplication $\times$ of context-free expressions.

\begin{prop}\label{prop:bisimcong}
If $s \sim t$ and $u \sim v$, then also $s + u \sim t + v$ and $s \times t \sim u \times v$.
\end{prop}

\proof
For the first claim, construct the relation
\[
R \colonequals \{ (s + u, t + v) \,|\, s, t, u, v \in \mathcal{T}_\mu \textrm{ and } s \sim t \textrm{ and } u \sim v \}.
\]
$R$ is a bisimulation, because if $(s + u, t + v) \in R$, then
\[
o(s + u) = o(s) + o(u) = o(t) + o(v) = o(t + v)
\]
and
\[
(s + u)_a = s_a + u_a \; R \; t_a + v_a = (t + v)_a.
\]
For the second claim, construct the relation
\[
R \colonequals \{ (s \times u, t \times v) \,|\, s, t, u, v \in \mathcal{T}_\mu \textrm{ and } s \sim t \textrm{ and } u \sim v \}.
\]
$R$ is a bisimulation up to $+$, because if $(s \times u, t \times v) \in R$, then
\[
o(s \times u) = o(s) \cdot o(u) = o(t) \cdot o(v) = o(t \times v)
\]
and
\[
(s \times u)_a = (s_a \times u) + (j(o(s)) \times u_a) \; R + R \; (t_a \times v) + (j(o(t)) \times v_a) = (t \times v)_a.\eqno{\qEd}
\]

\noindent Moreover, we can combine the notions of bisimulation up to $+$ and bisimulation up to bisimlarity, yielding the following `bisimulation up to lemma', which we will present without proof:

\begin{lem}
If $R$ is a relation on $\mathcal{T}_\mu$, such that whenever $(r_1, r_2) \in R$, we have $o(r_1) = o(r_2)$ and for all $a \in A$ there are $s, t \in \mathcal{T}_\mu$, such that $(r_1)_a \sim s$, $(r_2)_a \sim t$, and either $s \; R \; t$, or $s \; R + R \; t$, or $s \; R + (R + R) \; t$, then $r_1 \sim r_2$.
\end{lem}

For a more comprehensive view of bisimulation up to, including its categorical ramifications, we refer to \cite{RoBoRu12}.

We now can establish that the set of context-free expressions modulo $\sim$ forms an idempotent semiring:

\begin{prop}\label{prop:bisims}
For all $s, t, u \in \mathcal{T}_\mu$, the following hold:
\[
\begin{array}{ccc}
\bar{0} + t \sim t& & s + t \sim t + s \\
t + \bar{0} \sim t & & s + (t + u) \sim (s + t) + u \\
\bar{0} \times t \sim \bar{0} & & \bar{1} \times t \sim t \\
t \times \bar{0} \sim \bar{0} & & t \times \bar{1} \sim t \\
s \times (t + u) \sim (s \times t) + (s \times u) & & s \times (t \times u) \sim (s \times t) \times u \\
(s + t) \times u \sim (s \times u) + (t \times u)
\end{array}
\]
\end{prop}

\proof
See Appendix.
\qed

These laws can be seen as a partial (sound but not complete) axiomatization of behavioural equivalence between context-free expressions. Note that, because language equivalence of context-free languages is not decidable (see e.g.~\cite{HoMoUl06}), there cannot be any complete finitary axiomatization of behavioural equivalence.

As an illustration of context-free expressions over the Boolean semiring $\mathbb{B}$, consider the expression $\mu x . (\bar{1} + (\bar{a} \times (x \times \bar{b})))$ which is mapped onto the language $\{ a^n b^n \}$. As another example, consider the following expression:
\[
\mu x. (\bar{1} + ((\bar{a} \times (x \times \bar{a})) + (\bar{b} \times (\mu y. (\bar{1} + ((\bar{a} \times \bar{0}) + (\bar{b} \times (y \times \bar{a})))) \times \bar{a})))).
\]\vspace{-4 pt}

\noindent In the next subsection, it will become clear that this expression corresponds to the language
\[
\{ a^mb^na^{m + n} \,|\, m, n \in \mathbb{N} \}
\]
from the earlier examples.

\subsection{From Systems of Equations to $\mu$-Expressions}

Assume we have a coalgebra generated by a syntactic system of behavioural differential equations $(o, \delta): X \to \mathbb{B} \times \mathcal{T}_K(X)^A$, and a term $t \in \mathcal{T}_K(X)$. From Section~\ref{sec:langeq}, we know that this term is mapped by the final homomorphism to a context-free language. In this section, we will look for a context-free expression, obtained by a process of repeated substitution, corresponding to this term in the sense that it is mapped onto the same language (or power series).

The correspondence between coalgebras and $\mu$-expressions can be proved in various ways. The techniques used here, moreover, have by themselves little to do with context-free languages themselves, and can be seen as an instance of a more general transformation between coalgebras and guarded $\mu$-expressions. In \cite{Mi10} , \cite{Si10} and \cite{SiBoRu10}, the equivalence between expressions and coalgebras is established through the notion of (uniform) syntactic substitutions, which differ from familiar uniform substitutions in the sense that no renaming of freely occurring variables takes place. In our presentation, instead of uniform syntactic substitutions, we will use the notion of \emph{single syntactic substitutions}, in which a variable occurring in an expression (whether freely or not) is replaced by a corresponding $\mu$-expression.

Formally, given a function $\phi$ mapping variables $x \in X$ to guarded terms $\phi(x)$ and an $x \in X$, a term $t' \in \mathcal{T}^-_\mu$ is a \emph{single syntactic substitution} of $t \in \mathcal{T}^-_\mu$ from $x$, whenever either
\begin{enumerate}[(1)]
\item $t = x$ and $t' = \mu x. \phi(x)$;
\item $t = s + u$ and $t' = s' + u$, where $s'$ is a single syntactic substitution of $s$ from $x$;
\item $t = u + s$ and $t' = u + s'$, where $s'$ is a single syntactic substitution of $s$ from $x$;
\item $t = s \times u$ and $t' = s' \times u$, where $s'$ is a single syntactic substitution of $s$ from $x$;
\item $t = u \times s$ and $t' = u \times s'$, where $s'$ is a single syntactic substitution of $s$ from $x$; or
\item $t = \mu y. u$, $t' = \mu y. u'$, where $u'$ is a single syntactic substitution of $u$ from $x$.
\end{enumerate}

Here, we emphasize that we have \emph{not} imposed any requirement about the variables being substituted occurring freely. The main reason why we can do without this usual requirement, in this situation, is the fact that our expressions are constructed in such a way that---and this fact we will soon prove using bisimulation---every variable corresponds to one class of terms, all of which are behaviourally equivalent. In other words, our proof technique will relate a term to \emph{any} closure of it, rather than just one specific, or canonical, closure. This not only significantly simplifies matters, but furthermore combines well with the inherently relational nature of bisimulations.

A \emph{chain of syntactic substitutions} (w.r.t.~an assignment $\phi: X \to \mathcal{T}^-_\gamma$) is a list of terms $t_0, \ldots, t_{n - 1}$ such that, for each $i \in \mathbb{N}$ with $i < n - 1$, $t_{i+1}$ is a single syntactic substitution of $t_i$ (again w.r.t. $\phi$). We call a term $t' \in \mathcal{T}^-_\gamma$ obtainable from another term $t \in \mathcal{T}^-_\gamma$ by a chain of syntactic substitutions whenever there is a natural number $n$ and a chain of syntactic substitutions $t_0, \ldots t_{n - 1}$ such that $t_0 = t$ and $t_{n - 1} = t'$.

The following lemma establishes some elementary results about chains of syntactic substitutions:

\begin{lem}\label{lem:chains}
A term $t'$ is obtainable from $t$ by a chain of syntactic substitutions with respect to an assignment $\phi$ iff exactly one of the following conditions holds:
\begin{enumerate}[\em(1)]
\item $t = t'$, and either $t = \bar{a}$ for some $a \in A$ or $t = \bar{k}$ for some $k \in K$.
\item $t = x$ for some $x \in X$, and either $t' = x$ or $t'$ is obtainable from $\mu x. \phi(x)$ by a chain of syntactic substitutions.
\item $t = u + v$, $t' = u' + v'$, and $u'$ and $v'$ are obtainable from $u$ and $v$ by chains of syntactic substitutions.
\item $t = u \times v$, $t' = u' \times v'$, and $u'$ and $v'$ are obtainable from $u$ and $v$ by chains of syntactic substitutions.
\item $t = \mu x. u$, $t' = \mu x. u'$, and $u'$ is obtainable from $u$ by a chain of syntactic substitutions.
\end{enumerate}
\end{lem}

\proof
Induction on the length of chains, making use of the definition of single syntactic substitutions.
\qed

We are especially interested in chains of syntactic substitutions, where the resulting term does not contain any free variables, or only a limited set of free variables. We say a term $t'$ is a $Z$-\emph{pseudoclosure} of $t$ for a set $Z \subseteq X$ of variables, with respect to a mapping $\phi: X \to \mathcal{T}^-_\gamma$, if $t'$ is obtainable by a chain of syntactic substitutions from $t$, and $t'$ only contains free variables from $Z$. We call a $\emptyset$-pseudoclosure simply a \emph{closure}.

Given a system of equations $(X, (o, \delta))$, we will associate to every variable $x$ the guarded $\mu$-expression
\[
\mathbf{mu}(x) \colonequals j(o(x)) + \sum_{a \in A} (\bar{a} \times x_a)
\]
and call it the \emph{corresponding} or \emph{associated} $\mu$-expression. In order to give the $\sum$-expression a unique interpretation, however, we first need to assume a total ordering on the alphabet $A$. However, because $A$ is finite, such an ordering always exists.

As a continuation of our running example, recall the system of equations corresponding to the language $\{ a^nb^ma^{n + m} \,|\, m, n \in \mathbb{N} \}$. From this system of equations, we obtain the following canonical assignment $\mathbf{mu}$:
\begin{eqnarray*}
\mathbf{mu}(x) & = & \bar{1} + ((\bar{a} \times (x \times \bar{a})) + (\bar{b} \times (y \times \bar{a}))) \\
\mathbf{mu}(y) & = & \bar{1} + ((\bar{a} \times \bar{0}) + (\bar{b} \times (y \times \bar{a})))
\end{eqnarray*}
From $x$, we now obtain
\[
\mu x. (\bar{1} + ((\bar{a} \times (x \times \bar{a})) + (\bar{b} \times (y \times \bar{a}))))
\]
by means of a single syntactic substitution, and another single syntactic substitution then gives us
\[
\mu x. (\bar{1} + ((\bar{a} \times (x \times \bar{a})) + (\bar{b} \times (\mu y. (\bar{1} + ((\bar{a} \times \bar{0}) + (\bar{b} \times (y \times \bar{a})))) \times \bar{a})))).
\]
This expression does not contain any free variables anymore, and therefore is a closure of $x$. However, because single syntactic substitutions are not restricted to free variables, we can still apply another single syntactic substitution to this expression, yielding
\[
\mu x. (\bar{1} + ((\bar{a} \times (\mu x. (\bar{1} + ((\bar{a} \times (x \times \bar{a})) + (\bar{b} \times (y \times \bar{a})))) \times \bar{a})) + (\bar{b} \times (\mu y. (\bar{1} + ((\bar{a} \times \bar{0}) + (\bar{b} \times (y \times \bar{a})))) \times \bar{a}))))
\]
as another closure of $x$.

Some general laws about closures and pseudoclosures are easily established. For the following lemma, $\mathbf{fv}(t)$ refers to the set of variables occurring freely in the expression $t$.

\begin{lem}\label{lem:closure}
If $u'$ is a $W$-pseudoclosure of $u$ and $v'$ a $W$-pseudoclosure of $v$, then $u' + v'$ is a $W$-pseudoclosure of $u + v$, $u' \times v'$ is a $W$-pseudoclosure of $u \times v$, and $\mu x. u'$ is a $W - \{ x \}$-pseudoclosure of $\mu x. u$. Furthermore, if $t = u + v$, and $t'$ is a $W$-pseudoclosure of $t$, then $t'$ is of the form $u' + v'$, where $u'$ is a $W$-pseudoclosure of $u$, and $v'$ is a $W$-pseudoclosure of $v'$. The last fact again holds if we replace $+$ by $\times$.
\end{lem}

\proof
Use Lemma~\ref{lem:chains} together with the facts that $\mathbf{fv}(u + v) = \mathbf{fv}(u \times v) = \mathbf{fv}(u) \cup \mathbf{fv}(v)$, and $\mathbf{fv}(\mu x.u) = \mathbf{fv}(u) - \{ x \}$.
\qed

Using the previous proposition, we can establish that, for every term $t$, a closure $t'$ exists with respect to the assignment $\mathbf{mu}$. It should be noted, though, that this $t'$ generally is not unique: for a term $t$, in general, many closures exist.

\begin{prop}
Given a term $t \in \mathcal{T}_K(X)$ (that is, a $\mu$-free term), a set of variables $Z \subseteq X$, and an assignment $\phi: X \to \mathcal{T}^-_\gamma$ of variables to $\mu$-expressions, there exists a $Z$-pseudoclosure $t'$ of $t$ with respect to this assignment.
\end{prop}

\proof

By induction on the size of $X - Z$. If $|(X - Z)| = 0$, then $Z = X$, and the result is trivial because every term is its own $X$-pseudoclosure as witnessed by the chain of syntactic substitutions $(t)$.

Now say $|(X - Z)| = n$, assuming that statement of the theorem holds for all $W$ with $|(X - W)| < n$. Taking an arbitrary $x \in X - Z$, we obtain $|X - (Z \cup \{ x \})| < n$, so by the inductive hypothesis any term $t$ has a $Z \cup \{ x \}$-pseudoclosure $t'$. We now use this fact to prove, using structural induction on terms, that all terms $t$ also have $Z$-pseudoclosures.

\begin{enumerate}[(1)]
\item For terms of the form $\bar{k}$ for $k \in K$ and $\bar{a}$ for $a \in A$, the result is trivial as these terms do not contain any free variables, and hence are their own $Z$-pseudoclosures.
\item For the variable $x$, we know that there must be a $Z \cup \{ x \}$-pseudoclosure $u$ of $\phi(x)$. But now $\mu x. u$ is a $Z$-pseudoclosure of $x$.
\item For variables $y \ne x$, assume that $u$ is a $Z \cup \{ x \}$-pseudoclosure of $\phi(y)$, and $v$ is a $Z \cup \{ x \}$-pseudoclosure of $\phi(x)$. Then $u\{ \mu x. v / x \}$ is a $Z$-pseudoclosure of $\phi(y)$. Here $\{ \mu x. v / x \}$ denotes syntactic substitution of $\mu x. v$ for $x$, in which no renaming of free variables takes place. It is moreover easy to see that any such substitution $u\{ t/x \}$, where $t$ is obtainable by a chain of syntactic substitutions from $x$, is again obtainable by a chain of syntactic substitutions from $u$, the length of which is precisely equal to the number of free occurrences of $x$ in $u$. As a consequence, it now follows that $\mu y.u \{ \mu x. v / x \}$ is a $Z$-pseudoclosure of $y$.
\item For terms $t = u + v$ or $t = u \times v$, the result follows from Lemma~\ref{lem:closure} and the inductive hypothesis. \qed
\end{enumerate}

\noindent With the next proposition we construct a bisimulation up to bisimilarity between a coalgebra generated by a system of equations and the coalgebra of closed context-free expressions, relating every term $t \in \mathcal{T}_K(X)$ to all terms $t' \in \mathcal{T}_\mu$ that are closures of it w.r.t.~the (`canonical') assignment $\mathbf{mu}$.

\begin{prop}\label{prop:closurebisim}
Given a system of equations $(X, (o, \delta))$ together with the mapping $\mathbf{mu}$ just defined, the relation
\[
R = \{ (t, t') \,|\, t \in \mathcal{T}_K(X), t' \in \mathcal{T}_\mu \textrm{ and } t' \textrm{ is a closure of } t \textrm{ (w.r.t.~ $\mathbf{mu}$)} \}
\]
is a bisimulation up to bisimilarity between $(\mathcal{T}_K(X), (\bar{o}, \bar{\delta}))$ and the coalgebra of closed context-free expressions.
\end{prop}

\proof
The proof proceeds by proving the following claim using structural induction on terms $t$: for any $t'$ with $t \; R \; t'$, we have $\bar{o}(t) = o(t')$, and for each alphabet symbol $a$, there is a $u$, such that $t_a \; R \; u$ and $u \sim t'_a$.

\begin{iteMize}{$\bullet$}
\item If $t = \bar{k}$ for some $k \in K$, then it follows that $t' = t$ (as the only chain of single syntactic substitutions starting from $t$ has length $0$), and hence $\bar{o}(t) = o(t')$, and $t_a = \bar{0} \; R \; \bar{0} \sim \bar{0} = t'_a$.
\item If $t = \bar{b}$ for some $b \in A$, then it follows that $t' = t$, and hence $\bar{o}(t) = \bar{0} = o(t')$. For any $a \in A$ with $a \ne b$, we have $t_a = \bar{0} \; R \; \bar{0} \sim \bar{0} = t'_a$, and furthermore $t_b = \bar{1} \; R \; \bar{1} \sim \bar{1} = t'_b$, completing the case.
\item If $t = x$ for some $x \in X$, it follows that $t'$ must be obtainable by a chain of syntactic substitutions from $\mathbf{mu}(x)$. Because
\[
\mathbf{mu}(x) = j(o(x)) + \sum_{a \in A} (\bar{a} \times x_a)
\]
it follows that $t'$ must be of the form
\[
\mu x.(j(o(x)) + \sum_{a \in A} (\bar{a} \times \psi(a)))
\]
with each $\psi(a)$ a $\{ x \}$-pseudoclosure of $x_a$. However, it now follows that each uniform substitution
\[
\psi(a)[t'/x]
\]
is a closure of $x_a$, as a result of the equality $\psi(a)[t'/x] = \psi(a)\{ t'/x \}$ and the fact that $t'$ is a closure of $x$.

We now get
\[
\bar{o}(t) = \bar{o}(x) = o(t')
\]
and
\[
t_a = x_a \; R \; \psi(a)[t'/x] \sim \left( (j(o(x)) + \sum_{a \in A} (\bar{a} \times \psi(a))[t'/x] \right)_a = t'_a
\]
completing the case.
\item If $t = s + u$, use the inductive hypothesis that the claim holds for $s$ and $u$. Hence, if $s \; R \; s'$, we have $\bar{o}(s) = o(s')$ and for each alphabet symbol $a$, there is a $v$ such that $s_a \; R \; v \sim s'_a$, and if $u \; R \; u'$, we have $\bar{o}(u) = o(u')$ and for each alphabet symbol $a$, there is a $w$ such that $u_a \; R \; w \sim u'_a$.

If $t \; R \; t'$, then it follows from Lemma~\ref{lem:closure} that $t'$ is of the form $s' + u'$, with $s'$ a closure of $s$ and $u'$ a closure of $u$. However, this implies $s \; R \; s'$ and $u \; R \; u'$ and the foregoing now gives $v$ and $w$ with $s_a \; R \; v \sim s'_a$ and $u_a \; R \; w \sim u'_a$. First, we now get $\bar{o}(t) = \bar{o}(s) + \bar{o}(u) = o(s') + o(u') = o(t')$. Moreover, again by Lemma~\ref{lem:closure}, $v + w$ is a closure of $t_a = s_a + u_a$, and by Proposition~\ref{prop:bisimcong}, $v + w \sim s'_a + u'_a = t'_a$, completing the case.
\item If $t = s \times u$, again use the inductive hypothesis that the claim holds for $s$ and $u$. Hence, if $s \; R \; s'$, we have $\bar{o}(s) = o(s')$ and for each alphabet symbol $a$, there is a $v$ such that $s_a \; R \; v \sim s'_a$, and if $u \; R \; u'$, we have $\bar{o}(u) = o(u')$ and for each alphabet symbol $a$, there is a $w$ such that $u_a \; R \; w \sim u'_a$.

If $t \; R \; t'$, then $t'$ is of the form $s' \times u'$ by Lemma~\ref{lem:closure}, with $s'$ a closure of $s$ and $u'$ a closure of $u$. This again implies $s \; R \; s'$ and $u \; R \; u'$ and the foregoing now gives $v$ and $w$ with $s_a \; R \; v \sim s'_a$ and $u_a \; R \; w \sim u'_a$. We now get
\[
\bar{o}(t) = \bar{o}(s) \cdot \bar{o}(u) = o(s') \cdot o(u') = o(t')
\]
and
\begin{eqnarray*}
t_a & = & (s_a \times u) + (j(\bar{o}(s)) \times u_a) \\
& R & (v \times u') + (j(o(s')) \times w) \\
& \sim & (s'_a \times u') + (j(o(s')) \times u'_a) \\
& = & t'_a
\end{eqnarray*}
completing the case.\qed
\end{iteMize}

\noindent Returning to our example, this proposition directly establishes that the expression
\[
\mu x. (1 + ((\bar{a} \times (x \times \bar{a})) + (\bar{b} \times (\mu y. (1 + ((\bar{a} \times \bar{0}) + (\bar{b} \times (y \times \bar{a})))) \times \bar{a})))).
\]
corresponds to the language
\[
\{ a^nb^ma^{m + n} \,|\, m, n \in \mathbb{N} \}
\]
as it was precisely obtained as a closure with respect to the canonical assignment $\mathbf{mu}$.

More generally, the two previous propositions combined directly imply that, for any term in a coalgebra generated by a system of equations (and, hence, for every context-free language), we can use any closure of it with respect to the canonical mapping $\mathbf{mu}$ as a bisimilar context-free expression. Hence for every context-free language we can find a context-free expression that is mapped to it by the final homomorphism:

\begin{thm}
Let $L$ be a context-free power series over a commutative semiring. There exists a context-free expression $t$ over the same semiring such that $\llbracket t \rrbracket = L$.
\end{thm}

\subsection{From Context-Free Expressions to Systems of Equations}
\label{sec:cfetosyseq}

Going in the other direction, the recipe is as follows: given a context-free expression $u$ in which every variable is bound by a $\mu$-operator just once, we `deconstruct' this expression into a system of equations, and a term $t$, of which a closure $t'$ exists with $t' \sim u$. Then Proposition~\ref{prop:closurebisim} directly gives us the result, that there is a system of equations $(X, (o, \delta))$, a variable $x \in X$, such that $t' \sim t$ with respect to the coalgebra of closed context-free expressions and the coalgebra $(\mathcal{T}_K(X), (\bar{o}, \bar{\delta}))$ generated by $(X, (o, \delta))$.  Hence, the final homomorphism maps $t$ to a context-free language.

By applying a process of $\alpha$-renaming, we can obtain an expression $t'$ from any expression $t$ such that, in $t'$, no variable is bound twice, or, in other words, such that there are no two distinct subexpressions of $t'$ that bind the same variable. We will state the following lemma without proof:

\begin{prop}
Given a closed $\mu$-expression $t$, there is a closed $\mu$-expression $t'$ such that no two subexpressions of $t'$ are $\mu$-expressions binding the same variable, with $t \sim t'$.
\end{prop}

We now define the notion of the \emph{$\mu$-pruning} $\mathbf{mp}(t)$ of terms $t \in \mathcal{T}^-_\mu$ inductively by setting
\[
\mathbf{mp}(\mu x.g) = x,
\]
and furthermore $\mathbf{mp}(t) = t$ whenever $t = \bar{a}$ for some $a \in A$, $t = \bar{k}$ for some $k \in K$, or $t = x$ for some $x \in X$; $\mathbf{mp}(t + u) = \mathbf{mp}(t) + \mathbf{mp}(u)$; and $\mathbf{mp}(t \times u) = \mathbf{mp}(t) \times \mathbf{mp}(u)$.

Now, given an expression $t \in \mathcal{T}_\mu$ such that no two subexpressions of $t$ bind the same variable, construct the assignment $\psi: X \to \mathcal{T}^-_\mu$, where $X$ is the set of variables occurring in $t$ (and hence, as $t$ is closed, equal to the set of variables bound by subexpressions of $t$), and for all $x \in X$,
\[
\psi(x) = \mathbf{mp}(u) \qquad \textrm{where $\mu x. u$ is the unique subexpression of $t$ binding $x$.}
\]
Moreover, note that for all $x \in X$, $\psi(x)$ is guarded, and that any expression $t$ such that no two subexpressions of $t$ bind the same variable is a closure of its $\mu$-pruning.

We now construct a syntactic system of behavioural differential equations on $X$. In order to be able to do this, we first define an output and derivative operations $O$ and $\Delta$ on guarded expressions inductively as follows:
\begin{center}
\begin{tabular}{c|c|c}
$t \in \mathcal{T}^-_\gamma$ & $O(t) \in K$ & $\Delta(t)(a) \in \mathcal{T}^-_\mu$ \\
\hline
\rule{0pt}{2.2ex}$\bar{k} \; (k \in K)$ & $k$ & $\bar{0}$ \\
$u + v$ & $o(u) + o(v)$ & $u_a+v_a$\\
$\bar{b} \times v \; (b \in A)$ & $0$ & $\textbf{if $b = a$ then $v$ else $\bar{0}$}$
\end{tabular}
\end{center}
and now set, for all $x \in X$, $o(x) = O(\psi(x))$ and $x_a = \Delta(\psi(x))(a)$.

\begin{prop}\label{prop:2ndclosure}
The relation
\[
R = \{ (t, t') \,|\, t \in \mathcal{T}_K(X), t' \in \mathcal{T}_\mu \textrm{ and } t' \textrm{ is a closure of } t \textrm{ (w.r.t.~ $\psi$)} \}
\]
is a bisimulation up to bisimilarity between $(\mathcal{T}_K(X), (\bar{o}, \bar{\delta}))$ (as just defined) and the coalgebra of closed context-free expressions.
\end{prop}

\proof
The proof again proceeds by proving the following claim using structural induction on terms $t$: for any $t'$ with $t \; R \; t'$, we have $\bar{o}(t) = o(t')$, and for each alphabet symbol $a$, there is a $u$, such that $t_a \; R \; u$ and $u \sim t'_a$.

\begin{iteMize}{$\bullet$}
\item If $t = x$ for some $x \in X$, it follows that $t'$ must be obtainable by a chain of syntactic substitutions from $\mu x . \psi(x)$, and must hence be of the form $\mu x. u$, where $u$ is a $\{ x \}$-pseudoclosure of $\psi(x)$, and hence $u[t'/x]$ is a closure of $\psi(x)$.

We now get
\[
t_a = x_a = \Delta(\psi(x))(a) \; R \; \Delta(u[t'/x])(a) \sim (u[t'/x])_a = (\mu x. u)_a = t'_a
\]
where $\Delta(\psi(x))(a)$ and $\Delta(u[t'/x])(a)$ are related by $R$, because it can be derived from the fact that $u[t'/x]$  is a closure of $\psi(x)$, that also $\Delta(u[t'/x])(a)$ is a closure of $\Delta(\psi(x))(a)$ (to verify this, check the definition of $\Delta$ and use structural induction on guarded terms). Likewise, using the earlier established bisimilarities it can be shown that for all closed and guarded terms $g$, $\Delta(g)(a) \sim g_a$.

Using a similar argument, we also get
\[
\bar{o}(t) = \bar{o}(x) = O(\psi(x)) = O(u[t'/x]) = o(u[t'/x]) = o(t')
\]
where again the 2nd and 3rd equality follow from induction on guarded terms.

\item For all the other cases, the argumentation from Proposition~\ref{prop:closurebisim} can be used without modification.\qed
\end{iteMize}

\noindent This proposition directly leads to the following theorem:

\begin{thm}
For all $t \in \mathcal{T}_\mu$, $\llbracket t \rrbracket$ is context-free (or, depending on the underlying semiring, constructively algebraic).
\end{thm}

\proof
Take an expression $t'$ with $t \sim t'$ such that no two subexpressions of $t'$ bind the same variable. Proposition~\ref{prop:2ndclosure} now gives a syntactic system of behavioural differential equations $(X, (o, \delta))$, such that $\mathbf{mp}(t) \; R \; t$, from which it directly follows that $\llbracket t \rrbracket$ is context-free.
\qed

\section{The generalized powerset construction}
\label{sec:category}

In~\cite{SiBoBoRu10}, a categorical generalization of the powerset construction, a well-known method to transform nondeterministic finite automata into deterministic automata, is presented. The finite powerset functor $\mathcal{P}_\omega$ here is generalized to an arbitrary monad $T$, and instead of determinizing nondeterministic automata into deterministic automata we now transform $FT$-coalgebras into $F$-coalgebras using a categorically analogous technique. The familiar powerset construction can then be seen as an instance of this general construction, where $F$ is the functor $\mathbb{B} \times (-)^A$ representing deterministic automata, and $T$ is the powerset monad $\mathcal{P}_\omega$.

In order to cast the work from this article in a more general categorical light, it would be useful to be able to see the constructions from the earlier sections as instances of this framework. Hence, in the remainder of this section, we will first recall some of the main results from \cite{SiBoBoRu10}, and then investigate what needs to be done in order to be able to see the constructions from Sections \ref{sec:cfgcoalgebra} and \ref{sec:langeq} as instances of the generalized powerset construction.

\subsection{The framework}

To start, we present the categorical formulation of the familiar powerset construction, for the powerset monad and deterministic automata. We start by fixing an alphabet $A$.

Say, we are given a nondeterministic automaton, $(o, \delta): X \to \mathbb{B} \times (\mathcal{P}_\omega (X))^A$. From this automaton, we will construct a new, deterministic, automaton $(\hat{o}, \hat{\delta}): \mathcal{P}_\omega (X) \to \mathbb{B} \times (\mathcal{P}_\omega (X))^A$, accepting the same language.

This can be done by specifying, for all $Y \in \mathcal{P}_\omega (X) $:
\[
\hat{o}(Y) = \bigvee_{y \in Y} o(y) \,, \quad \textrm{and} \quad Y_a = \bigcup_{y \in Y} y_a
\]
This extension enables us to obtain an interpretation of the original nondeterministic automaton by means of the final homomorphism $\llbracket - \rrbracket$, as in the following commuting diagram:
\begin{diagram}
X & \rTo_{\{ - \}} & \mathcal{P}_\omega (X) & \rDotsto_{\llbracket - \rrbracket} & \mathcal{P}(A^*) \\
\dTo_{(o, \delta)} & \ldTo_{(\hat{o}, \hat{\delta})} & & & \dTo \\
\mathbb{B} \times \mathcal{P}_\omega(X)^A & & \rDotsto & & \mathbb{B} \times \mathcal{P}(A^*)^A
\end{diagram}

Another way to look at this construction, is to observe that we can assign a $\mathcal{P}_\omega$-algebra structure to the set $\mathbb{B} \times \mathcal{P}_\omega (X)^A$, such that there is a unique $\mathcal{P}_\omega$-algebra homomorphism from $\mathcal{P}_\omega (X)$, the free $\mathcal{P}_\omega$-algebra over $X$, to $\mathbb{B} \times \mathcal{P}_\omega (X)^A$, making the triangle in the above diagram commute.

We can now formulate the general case as follows: let $T$ be a monad on a category $\mathbf{C}$ and let $F$ be an endofunctor on $\mathbf{C}$, and consider any coalgebra $c: X \to FTX$. The aim now is to extend $c$ into some $\hat{c}: TX \to FTX$, making the diagram
\begin{diagram}
X & \rTo_{\eta_X} & TX \\
\dTo_c & \ldTo_{\hat{c}} & \\
FTX & &
\end{diagram}
commute. Here $\eta$ is the unit of the monad $T$.

This extension can be constructed by assigning an appropriate $T$-algebra structure $(FTX, \alpha)$ to $FTX$. We will now have a unique $T$-algebra morphism
\[
\hat{c}: (TX, \mu_X) \to (FTX, \alpha)
\]
from the free algebra $(TX, \mu_X)$ into $(FTX, \alpha)$ making the above diagram commute.

When $F$ has a final coalgebra $(\Omega, \omega)$, the diagram can be extended as follows:
\begin{diagram}
X & \rTo_{\eta_X} & TX & \rDotsto_{\llbracket - \rrbracket} & \Omega \\
\dTo_c & \ldTo_{\hat{c}} & & & \dTo_\omega \\
FTX & & \rDotsto & & F \Omega
\end{diagram}
We will later see that the main constructions in this article can all be seen as instances of this generalized powerset construction.

As a final remark, note that if $\lambda: TF \to FT$ is a distributive law of the monad $T$ over $F$, then the required $T$-algebra structure $FTX$ can be canonically obtained as follows:
\begin{diagram}
TFTX & \rTo^{\lambda_{TX}} & FTTX & \rTo^{F\mu_X} & FTX
\end{diagram}
This algebra structure is consistent with the algebra structure one can give (by finality) on the final coalgebra $(\Omega, \omega)$ in the sense that the coalgebra map $\llbracket - \rrbracket$ is a also a $T$-algebra homomorphism.
(Earlier in this article, we obtained the same compositionality results directly without reliance on distributive laws or other categorical methods.)
Here, like in the case of the generalized powerset construction, again some subtleties arise and the easiest approach is to define a distributive law of the monad $\mathcal{T}$ over the copointed functor $- \times (\mathbb{B} \times (-)^A)$. From this we can obtain a distributive law of the monad $\mathcal{P}_\omega(-^*)$ over $- \times (\mathbb{B} \times (-)^A)$, as presented in \cite{BoHaKuRo13}, to which we refer for further details. For a more general background on distributive laws, we refer to e.g.~\cite{Ba04} or \cite{Kl11}.

\subsection{The constructions from this article as an instance of the framework}

It is also possible to regard the constructions from Sections~\ref{sec:cfgcoalgebra} and \ref{sec:langeq} as instances of the generalized powerset construction presented above.

In order to be able to see the extension from $\mathbb{B} \times \mathcal{P}_\omega((-)^*)^A$-coalgebras $(o, \delta)$ into $\mathbb{B} \times (-)^A$-coalgebras $(\hat{o}, \hat{\delta})$ as an instance of the generalized powerset construction, we would need to provide an appropriate algebra structure on $\mathbb{B} \times \mathcal{P}_\omega(X^*)^A$. First observe that $\mathcal{P}_\omega(X^*)$ is the free idempotent semiring on $X$ or, in other words: given any idempotent semiring $K$ and a function $f: X \to K$, there is a unique semiring morphism $\hat{f}: \mathcal{P}_\omega(X^*) \to K$, such that $\hat{f} \circ \eta_X = f$. Hence, it is enough to assign an idempotent semiring structure to $\mathbb{B} \times \mathcal{P}_\omega(X^*)^A$ in order to obtain $(\hat{o}, \hat{\delta})$ via a unique mapping property. Although we already know that the diagram
\begin{diagram}
X & \rTo_{\eta_X} & \mathcal{P}_\omega(X^*) \\
\dTo_{(o, \delta)} & \ldTo_{(\hat{o}, \hat{\delta})} & \\
\mathbb{B} \times \mathcal{P}_\omega(X^*)^A
\end{diagram}
commutes in $\mathbf{Set}$, we would like $(\hat{o}, \hat{\delta})$ to be a semiring morphism, in order to be able to see this construction as an instance of the generalized powerset construction.

However, this cannot be done right away: we first need to make a suitable extension, from $\mathbb{B} \times \mathcal{P}_\omega((-)^*)^A$-coalgebras to $\mathbb{B} \times \mathcal{P}_\omega((- + A)^*)^A$-coalgebras, which can be seen as corresponding to grammars $(X, p)$ in \emph{weak Greibach normal form}, where for each $x \in X$, $p(x) \subseteq A (X + A)^* \cup \{ \epsilon \}$ (rather than $p(x) \subseteq AX^* \cup \{ \epsilon \}$ as in the case of the regular Greibach normal form). The appropriate diagram now is
\begin{diagram}
X & \rTo_{\eta_X} & \mathcal{P}_\omega((X + A)^*) \\
\dTo_{(o, \delta)} & \ldTo_{(\hat{o}, \hat{\delta})} & \\
\mathbb{B} \times \mathcal{P}_\omega((X + A)^*)^A
\end{diagram}
and in order to give a full inductive specification of $(\hat{o}, \hat{\delta})$, we need to extend the earlier specification with the equations $\hat{o}(b) = 0$ and $b_a = i((b = a)\textbf{?})$ for all $a, b \in A$. Moreover, we need to change the definition of the derivative of products $\{ xs \}$ for $x \in X$ and $k \in K$ by decomposing the occurrence of $s$ in the derivative into $i(\hat{o}(s)) \cup \left( \bigcup_{b \in A} \{ b \} \{s\}_b \right)$, yielding
\begin{center}
\begin{tabular}{c|c|c}
$S$ & $\hat{o}(S)$ & $S_a\, (a \in A)$ \\
\hline $\{ \epsilon \}$ & $1$ & $\emptyset$ \\
\specialcell{ $\{ x s \}$ \\ $(x \in X, s \in (A + X)^*)$} & $o(x) \land \hat{o}(\{ s \})$ & $x_a \cdot \left( i(\hat{o}(s)) \cup \left( \bigcup_{b \in A} \{ b \} \{ s \}_b \right) \right) \cup i(o(x)) \{ s \}_a$ \\
\specialcell{ $\{ b s \}$ \\ $(b \in A, s \in (A + X)^*)$} & 0 & $i((b = a)\textbf{?}) \cdot \left(i(\hat{o}(s)) \cup \left( \bigcup_{c \in A} \{ c \} \{ s \}_c \right) \right)$ \\
\specialcell{$\bigcup_{i \le n} \{ s_i \}$ \\ $(n \in \mathbb{N}, w_i \in X^*)$} & $\bigvee_{i \le n} \hat{o}(\{ w_i \})$ & $\bigcup_{i \le n} \{ s_i \}_a$
\end{tabular}
\end{center}
as the behavioural specification of the $\mathbb{B} \times (-)^A$-coalgebra structure on $\mathcal{P}_\omega((X + A)^*)$.

Furthermore, we can now indeed define an idempotent semiring structure on $\mathbb{B} \times \mathcal{P}_\omega((X + A)^*)^A$. Because $\mathcal{P}_\omega((X + A)^*)$ is the free idempotent semiring on $X + A$, we need to give an interpretation of nonterminals $x \in X$ as well as alphabet symbols $a \in A$. Constants, addition, and multiplication now can be specified as follows:
\begin{center}
\begin{tabular}{c|c}
$\mathtt{0}$ & $(0, \lambda a.\emptyset)$ \\
$\mathtt{1}$ & $(1, \lambda a.\emptyset)$ \\
$x \in X$ & $(o(x), \lambda a.x_a)$ \\
$b \in A$ & $(0, \lambda a.i((b = a)\textbf{?}))$ \\
$(o_1, \delta_1) \oplus (o_2, \delta_2)$ & $(o_1 \lor o_2, \lambda a.(\delta_1(a) \cup \delta_2(a)))$ \\
$(o_1, \delta_1) \otimes (o_2, \delta_2)$ & $(o_1 \land o_2, \lambda a.( \delta_1(a) ( i(o_2) \cup (\bigcup_{b \in A} \{ b \} \delta_2(b))) \cup i(o_1) \delta_2(a)))$
\end{tabular}
\end{center}
We, however, still need to verify that this structure is indeed an idempotent semiring:
\begin{prop}\label{prop:hell}
$(\mathbb{B} \times \mathcal{P}_\omega((X + A)^*)^A, \otimes, \oplus, \mathtt{1}, \mathtt{0})$ is an idempotent semiring.
\end{prop}

\proof See Appendix. \qed

This semiring structure yields a unique semiring morphism $(O, \Delta)$ from $\mathcal{P}_\omega((X + A)^*)$, to $\mathbb{B} \times \mathcal{P}_\omega((X + A)^*)^A$ compatible with $(o, \delta)$, and can be combined with the unique homomorphism into the final coalgebra in the following commuting diagram:
\begin{diagram}
X & \rTo_{\eta_X} & \mathcal{P}_\omega((X + A)^*) & \rDotsto_{\llbracket - \rrbracket} & \mathcal{P}(A^*) \\
\dTo_{(o, \delta)} & \ldTo_{(O, \Delta)} & & & \dTo \\
\mathbb{B} \times \mathcal{P}_\omega((X + A)^*)^A & & \rDotsto & & \mathbb{B} \times \mathcal{P}(A^*)^A
\end{diagram}
However, in order to be certain that $(O, \Delta) = (\hat{o}, \hat{\delta})$, we either need to prove that $(\hat{o}, \hat{\delta})$ is a semiring morphism compatible with $(o, \delta)$, i.e.~that $(\hat{o}, \hat{\delta})(ST) = (\hat{o}, \hat{\delta})(S) \otimes (\hat{o}, \hat{\delta})(T)$ and $(\hat{o}, \hat{\delta})(S \cup T) = (\hat{o}, \hat{\delta})(S) \oplus (\hat{o}, \hat{\delta})(T)$ for all $S, T \in \mathcal{P}_\omega((X + A)^*)$, giving $(O, \Delta) = (\hat{o}, \hat{\delta})$ by the above unique mapping property, or we need to show this equivalence directly. Taking the latter route, we obtain the following proposition:

\begin{prop}
$(O, \Delta) = (\hat{o}, \hat{\delta})$ (w.r.t. the $\mathbb{B} \times (-)^A$-coalgebra structure on $\mathcal{P}_\omega((X + A)^*)$).
\end{prop}

\proof
We first show that for all words $s \in (X + A)^*$, $(O, \Delta)(\{ s \}) = (\hat{o}(\{ s \}), \hat{\delta}(\{ s \}))$ by induction on the length of $s$. Note that in what follows we will make use of the already established fact that $(O, \Delta)$ is a semiring morphism.

If $|s| = 0$, then $s = \epsilon$ and
\[
(O, \Delta)(\{ \epsilon \}) = \mathtt{1} = (1, \lambda a.\emptyset) = (\hat{o}(\{ \epsilon \}), \hat{\delta}(\{ \epsilon \}))
\]
so the property holds.

If $|s| > 0$, then either $s = xt$ for $x \in X$ and $t \in (X + A)^*$, or $s = at$ for $a \in A$ and $t \in (X + A)^*$. We now can use the inductive hypothesis that $(O, \Delta)(\{ t \}) = (\hat{o}(\{ t \}), \hat{\delta}(\{ t \}))$:

If $s = xt$ for $x \in X$ and $t \in (X + A)^*$, observe
\begin{equation*}\begin{split}
& (O, \Delta)(\{ xt \}) \\
=\; & (O, \Delta)(\{ x \}) \otimes (O, \Delta)(\{ t \}) \\
=\; & (o(x), \delta(x)) \otimes (\hat{o}(\{ t \}, \hat{\delta}(\{ t \})) \\
=\; & (o(x) \land \hat{o}(\{ t \}), \lambda a. (x_a \left[ i(\hat{o}(\{ t \})) \cup \bigcup_{b \in A} \{ b \} \{ t \}_b \right] \cup i(o(x))\{ t \}_a ))\\
=\; & (\hat{o}(\{ xt \}), \lambda a.\{ xt \}_a) \\
=\; & (\hat{o}(\{ xt \}), \hat{\delta}(\{ xt \}))
\end{split}\end{equation*}
and if $s = bt$ for $b \in A$ and $t \in (X + A)^*$, observe
\begin{equation*}\begin{split}
& (O, \Delta)(\{ bt \}) \\
=\; & (O, \Delta)(\{ b \}) \otimes (O, \Delta)(\{ t \}) \\
=\; & (0, \lambda a.i((b=a)\textbf{?})) \otimes (\hat{o}(\{ t \}, \hat{\delta}(\{ t \})) \\
=\; & (0 \land \hat{o}(\{ t \}), \lambda a. (i((b = a)\textbf{?}) \left[ i(\hat{o}(\{ t \})) \cup \bigcup_{c \in A} \{ c \} \{ t \}_c \right] \cup i(0)\{ t \}_a ))\\
=\; & (0, \lambda a. (i((b = a)\textbf{?}) \left[ i(\hat{o}(\{ t \})) \cup \bigcup_{c \in A} \{ c \} \{ t \}_c \right]))\\
=\; & (\hat{o}(\{ bt \}), \lambda a.\{ bt \}_a) \\
=\; & (\hat{o}(\{ bt \}), \hat{\delta}(\{ bt \})).
\end{split}\end{equation*}
Now let $S \in \mathcal{P}_\omega((X + A)^*)$ and observe
\begin{equation*}\begin{split}
& (O, \Delta)(S)
= (O, \Delta)(\bigcup_{s \in S} \{ s \})
= \bigoplus_{s \in S}(O, \Delta)(\{ s \})
= \bigoplus_{s \in S}(\hat{o}(\{ s \}), \hat{\delta}(\{ s \})) \\
=\; & \left( \bigvee_{s \in S}\hat{o}(s), \lambda a. \left[ \bigcup_{s \in S} \{ s \}_a \right] \right)
= (\hat{o}(S), \hat{\delta}(S)),
\end{split}\end{equation*}
completing the proof.
\qed

The construction in Section~\ref{sec:langeq}, too, can be seen as an instance of the generalized powerset construction. This time we need to provide a $\mathcal{T}$-algebra structure for $\mathbb{B} \times \mathcal{T}(X)^A$, which can be done as follows:
\begin{center}
\begin{tabular}{c|c}
$\bar{0}$ & $(0, \lambda a.\bar{0})$ \\
$\bar{1}$ & $(1, \lambda a.\bar{0})$ \\
$\bar{a}$ & $(0, \lambda b.j((b = a)\textbf{?}))$ \\
$(o_1, \delta_1) + (o_2, \delta_2)$ & $((o_1 \lor o_2), \lambda a.(\delta_1(a) + \delta_2(a)))$ \\
$(o_1, \delta_1) \times (o_2, \delta_2)$ & $((o_1 \land o_2), \lambda a.((j(o_1) \times \delta_2(a)) + (\delta_1(a) \times \sum_{b \in A} (\bar{b} \times \delta_2(a)))))$
\end{tabular}
\end{center}
As before, in order to make the expression $\sum_{b \in A} (\bar{b} \times \delta_2(a))$ precise, we again assume a canonical ordering on the alphabet $A$.

This $\mathcal{T}$-algebra structure on $\mathbb{B} \times \mathcal{T}(X)^A$ yields a unique $\mathcal{T}$-algebra morphism from the free $\mathcal{T}$-algebra $\mathcal{T}(X)$ to $\mathbb{B} \times \mathcal{T}(X)^A$ compatible with $(o, \delta)$. Moreover, this unique algebra morphism is behaviourally equivalent to the mapping $(\bar{o}, \bar{\delta})$ as defined before.

\section{Discussion}
\label{sec:discussion}

Our coalgebraic account of context-free languages in terms of grammar coalgebras and automata, syntactic systems of behavioural equations, and closed $\mu$-expressions can be taken as a starting point for a generalization in at least two different and orthogonal directions. In one direction (in which the first steps have already been taken, see Section~\ref{sec:generalize} and \cite{BoRuWi12}), we can generalize from languages towards power series in commutative semirings. Here, a good deal of work remains to be done in order to see how existing results from the literature on formal power series and languages can be cast into the coalgebraic framework, and to find uses of coalgebraic and coinductive techniques in this area.

In another direction, we can consider other languages of expressions for the functor $\mathbb{B} \times (-)^A$ to obtain different classes of languages. As an interesting example of this type, one could consider syntactic systems of behavioural differential equations for which the term at the right of each equation stems from the language of expressions:
\begin{eqnarray*}
t & \coloncolonequals & \bar{0} \,|\, \bar{1} \,|\, \bar{x} \, (x \in X) \,\,|\, t + t \,|\, \bar{a} \times t \, (a \in A)
\end{eqnarray*}

The semantic solution of such a system is given by regular languages, while the syntactic one is given by a language of expressions as studied in~\cite{SiBoRu10}. The corresponding notion of grammars for regular languages is then given by considering productions of the form $p: X \to \mathbb{B} \times \mathcal{P}_\omega(A^* \times X)^A$, i.e. \emph{right-linear grammars}~\cite{HoMoUl06}. When we generalize this language of expressions to the specification
\begin{eqnarray*}
t & \coloncolonequals & \bar{k} \, (k \in K) \,|\, \bar{x} \, (x \in X) \,|\, t + t \,|\, \bar{a} \times t \, (a \in A)
\end{eqnarray*}
over an arbitrary semiring $K$, we obtain the rational power series over $K$.  In both cases, it is also possible to give a corresponding notion of $\mu$-expressions.

It might also be worthwhile to look at systems of equations in which other coinductively specified operators than sum and concatenation (or, in the general case, the convolution product) are involved: we can for example investigate systems of equations where, on the right-hand side, the operator for language intersection (or, more generally, the Hadamard product) is allowed. Because the class of context-free languages is not closed under intersection, it is directly clear that adding this operator to the language of terms will yield a larger class of languages.

Further research directions include a coalgebraic characterization of context-free languages in terms of pushdown automata~\cite{HoMoUl06}, and the study of coinductive decision procedures for bisimilarity of subclasses of the context-free languages of which equivalence is decidable, such as deterministic pushdown automata~\cite{St01} and visibly pushdown automata~\cite{AlMa04}.

\section*{Acknowledgements}

We would like to thank Alexandra Silva and Jurriaan Rot, for valuable suggestions and discussions. Furthermore, we would like to thank the anonymous referees for their numerous valuable corrections and suggestions for improvements.

\appendix

\section*{Additional proofs}

\proof (of Proposition~\ref{prop:bisims})

For each of the relations, we construct a bisimulation (up to) establishing the claimed bisimilarity.

\begin{iteMize}{$\bullet$}
\item $0 + t \sim t$ and $t + 0 \sim t$:

Define
\[
R \colonequals \{ (0 + t, t) \,|\, t \in \mathcal{T}_\mu \} \cup \{ (t + 0, t) \,|\, t \in \mathcal{T}_\mu \}.
\]

If $(r_1, r_2) \in R$, then either $r_1 = 0 + t$ and $r_2 = t$ for some $t \in \mathcal{T}_\mu$, or $r_1 = t + 0$ and $r_2 = t$ for some $t \in \mathcal{T}_\mu$. If $(r_1, r_2) = (0 + t, t)$, then $o(0 + t) = o(0) \lor o(t) = o(t)$ and $(0 + t)_a = 0_a + t_a = 0 + t_a \; R \; t_a$, so $((0 + t)_a, t_a) \in R$. If $(r_1, r_2) = (t + 0, t)$, then $o(t + 0) = o(t) \lor o(0) = o(t)$ and $(t + 0)_a = t_a + 0_a \; R \; t_a$, so $((t + 0)_a, t_a) \in R$. Hence, $R$ is a bisimulation.\medskip

\item $s + t \sim t + s$:

Define
\[
R \colonequals \{ (s + t, t + s) \,|\, s, t \in \mathcal{T}_\mu \}
\]

If $(s + t, t + s) \in R$, then $o(s + t) = o(s) \lor o(t) = o(t) \lor o(s) = o(t + s)$ and
\[
(s + t)_a = s_a + t_a \; R \; t_a + s_a = (t + s)_a
\]
so $R$ is a bisimulation.\medskip

\item $s + (t + u) \sim (s + t) + u$:

Define
\[
R \colonequals \{ s + (t + u), (s + t) + u \,|\, s, t, u \in \mathcal{T}_\mu \}
\]

If $(s + (t + u), (s + t) + u) \in R$, then
\[
o(s + (t + u)) = o(s) \lor (o(t) \lor o(u)) = (o(s) \lor o(t)) \lor o(u) = o((s + t) + u)
\]
and
\[
(s + (t + u))_a = s_a + (t_a + u_a) \; R \; (s_a + t_a) + u_a = ((s + t) + u)_a
\]
so $R$ is a bisimulation.\medskip

\item $0 \times t \sim 0$ and $t \times 0 \sim 0$:

Define
\[
R \colonequals \{ (0 \times t, 0) \,|\, t \in \mathcal{T}_\mu \} \cup \{ (t \times 0, 0) \,|\, t \in \mathcal{T}_\mu \}
\]

If $(r_1, r_2) \in R$, then either $r_1 = 0 \times t$ and $r_2 = t$ for some $t \in \mathcal{T}_\mu$, or $r_1 = t \times 0$ and $r_2 = 0$ for some $t \in \mathcal{T}_\mu$. If $(r_1, r_2) = (0 \times t, 0)$, then $o(0 \times t) = o(0) \land o(t) = o(0)$ and
\[
(0 \times t)_a = (0_a \times t) + (i(o(0)) \times t_a) = (0 \times t) + (0 \times t_a) \; R + R \; 0 + 0 \sim 0 = 0_a,
\]
and if $(r_1, r_2) = (t \times 0, 0)$, then $o(t \times 0) = o(t) \land o(0) = o(0)$ and
\[
(t \times 0)_a = (t_a \times 0) + (i(o(t)) \times 0_a) = (t_a \times 0) + (i(o(t)) \times 0) \; R + R \; 0 + 0 \sim 0 = 0_a,
\]
so $R$ is a bisimulation up to $+$ and bisimilarity.\medskip

\item $1 \times t \sim t$ and $t \times 1 \sim t$:

Define
\[
R \colonequals \{ (1 \times t, t) \,|\, t \in \mathcal{T}_\mu \} \cup \{ (t \times 1, t) \,|\, t \in \mathcal{T}_\mu \}
\]

If $(r_1, r_2) \in R$, then either $r_1 = 1 \times t$ and $r_2 = t$ for some $t \in \mathcal{T}_\mu$, or $r_1 = t \times 1$ and $r_2 = t$ for some $t \in \mathcal{T}_\mu$. If $(r_1, r_2) = (1 \times t, t)$, then $o(1 \times t) = o(1) \land o(t) = o(t)$ and
\[
(1 \times t)_a = (1_a \times t) + (i(o(1)) \times t_a) = (0 \times t) + (1 \times t_a) \sim 0 + (1 \times t_a) \sim 1 \times t_a \; R \; t_a,
\]
and if $(r_1, r_2) = (t \times 1, t)$, then $o(t \times 1) = o(t) \land o(1) = o(t)$ and
\[
(t \times 1)_a = (t_a \times 1) + (i(o(t)) \times 1_a) = (t_a \times 1) + (i(o(t)) \times 0) \sim t_a \times 1 + 0 \sim t_a \times 1 \; R \; t_a,
\]
so $R$ is a bisimulation up to $+$ and bisimilarity.\medskip

\item $s \times (t + u) \sim (s \times t) + (s \times u)$.

Define
\[
R \colonequals \{ s \times (t + u), (s \times t) + (s \times u) \,|\, s, t, u \in \mathcal{T}_\mu \}
\]
If $(s \times (t + u), (s \times t) + (s \times u)) \in R$, then
\[
o(s \times (t + u)) = o(s) \land (o(t) \lor o(u)) = (o(s) \land o(t)) \lor (o(s) \land o(u)) = o((s \times t) + (s \times u))
\]
and
\begin{eqnarray*}
& & (s \times (t + u))_a \\
& = & (s_a \times (t + u)) + (i(o(s)) \times (t + u)_a) \\
& = & (s_a \times (t + u)) + (i(o(s)) \times (t_a + u_a)) \\
& R + R & (s_a \times t + s_a \times u) + ((i(o(s)) \times t_a) + (i(o(s)) \times u_a)) \\
& \sim & ((s_a \times t) + (i(o(s)) \times t_a)) + ((s_a \times u) + (i(o(s)) \times u_a)) \\
& = & ((s \times t) + (s \times u))_a
\end{eqnarray*}
so $R$ is a bisimulation up to $+$ and bisimilarity.\medskip

\item $(s + t) \times u \sim (s \times u) + (t \times u)$
Define
\[
R \colonequals \{ (s + t) \times u, (s \times u) + (t \times u) \,|\, s, t, u \in \mathcal{T}_\mu \}
\]
If $((s + t) \times u, (s \times u) + (t \times u)) \in R$, then
\[
o((s + t) \times u) = (o(s) \lor o(t)) \land o(u) = (o(s) \land o(u)) \lor (o(t) \land o(u)) = o((s \times u) + (t \times u))
\]
and
\begin{eqnarray*}
& & ((s + t) \times u)_a \\
& = & ((s + t)_a \times u) + (i(o(s + t)) \times u_a) \\
& = & ((s_a + t_a) \times u) + (i(o(s + t)) \times u_a) \\
& \sim & ((s_a + t_a) \times u) + ((i(o(s)) + i(o(t))) \times u_a \\
& R + R & ((s_a \times u) + (t_a \times u)) + ((i(o(s)) \times u_a) + (i(o(t)) \times u_a)) \\
& \sim & ((s_a \times u) + (i(o(s)) \times u_a)) + ((t_a \times u) + (i(o(t)) \times u_a)) \\
& = & ((s \times u) + (t \times u))_a
\end{eqnarray*}
so $R$ is a bisimulation up to $+$ and bisimilarity.\medskip

\item $s \times (t \times u) \sim (s \times t) \times u$
Define
\[
R \colonequals \{ s \times (t \times u), (s \times t) \times u \,|\, s, t, u \in \mathcal{T}_\mu \}
\]
If $(s \times (t \times u), (s \times t) \times u) \in R$, then
\[
o(s \times (t \times u)) = o(s) \land (o(t) \land o(u)) = (o(s) \land o(t)) \land o(u) = o((s \times t) \times u)
\]
and
\begin{eqnarray*}
& & (s \times (t \times u))_a \\
& = & (s_a \times (t \times u)) + (i(o(s)) \times (t \times u)_a)) \\
& = & (s_a \times (t \times u)) + (i(o(s)) \times ((t_a \times u) + (i(o(t)) \times u_a))) \\
& \sim & (s_a \times (t \times u)) + ((i(o(s)) \times (t_a \times u)) + (i(o(s)) \times (i(o(t)) \times u_a))) \\
& R + (R + R) & ((s_a \times t) \times u) + (((i(o(s)) \times t_a) \times u) + ((i(o(s)) \times i(o(t))) \times u_a)) \\
& \sim & (((s_a \times t) + (o(s) \times t_a)) \times u) + i(o(s \times t)) \times u_a) \\
& = & ((s \times t)_a \times u) + i(o(s \times t)) \times u_a) \\
& = & ((s \times t) \times u)_a
\end{eqnarray*}
so $R$ is a bisimulation up to $+$ and bisimilarity.\qed
\end{iteMize}\vspace{3 pt}

\proof (of Proposition~\ref{prop:hell})
We verify that the semiring axioms hold, as well as the axiom of idempotence.\vspace{3 pt}

\begin{iteMize}{$\bullet$}
\item $\mathtt{0} \oplus (o, \delta) = (o, \delta)$:
\[
\mathtt{0} \oplus (o, \delta)
= (0, \lambda a. \emptyset) \oplus (o, \delta)
= (0 \lor o, \lambda a. ((\lambda b. \emptyset)(a) \cup \delta(a)))
= (o, \lambda a.(\emptyset \cup \delta(a)))
= (o, \delta)
\]\vspace{-6 pt}

\item $(o, \delta) \oplus \mathtt{0} = (o, \delta)$:
\[
(o, \delta) \oplus \mathtt{0}
= (o, \delta) \oplus (0, \lambda a. \emptyset)
= (o \lor 0, \lambda a. (\delta(a) \cup (\lambda b. \emptyset)(a))
= (o, \lambda a. (\delta(a) \cup \emptyset))
= (o, \delta)
\]\vspace{-6 pt}

\item $(o_1, \delta_1) \oplus (o_2, \delta_2) = (o_2, \delta_2) \oplus (o_1, \delta_1)$:
\[
(o_1, \delta_1) \oplus (o_2, \delta_2)
= (o_1 \lor o_2, \lambda a. (\delta_1(a) \cup \delta_2(a)))
= (o_2 \lor o_1, \lambda a. (\delta_2(a) \cup \delta_1(a)))
= (o_2, \delta_2) \oplus (o_1, \delta_1)
\]\vspace{-6 pt}

\item $((o_1, \delta_1) \oplus (o_2, \delta_2)) \oplus (o_3, \delta_3) = (o_1, \delta_1) \oplus ((o_2, \delta_2) \oplus (o_3, \delta_3))$:
\begin{equation*}\begin{split}
& ((o_1, \delta_1) \oplus (o_2, \delta_2)) \oplus (o_3, \delta_3) \\
= \; & (o_1 \lor o_2, \lambda a. (\delta_1(a) \cup \delta_2(a))) \oplus (o_3, \delta_3) \\
= \; & ((o_1 \lor o_2) \lor o_3, \lambda a. ((\lambda b. (\delta_1(b) \cup \delta_2(b)))(a) \cup \delta_3(a))) \\
= \; & ((o_1 \lor o_2) \lor o_3, \lambda a. ((\delta_1(a) \cup \delta_2(a)) \cup \delta_3(a))) \\
= \; & (o_1 \lor (o_2 \lor o_3), \lambda a. (\delta_1(a) \cup (\delta_2(a) \cup \delta_3(a)))) \\
= \; & (o_1 \lor (o_2 \lor o_3), \lambda a. (\delta_1(a) \cup (\lambda b. (\delta_2(b) \cup \delta_3(b)))(a))) \\
= \; & (o_1, \delta_1) \oplus (o_2 \lor o_3, \lambda a. (\delta_2(a) \cup \delta_3(a))) \\
= \; & (o_1, \delta_1) \oplus ((o_2, \delta_2) \oplus (o_3, \delta_3))
\end{split}\end{equation*}\vspace{3 pt}

\item $\mathtt{1} \otimes (o, \delta) = (o, \delta)$:
\begin{equation*}\begin{split}
& \mathtt{1} \otimes (o, \delta) \\
= \; & (1, \lambda a. \emptyset) \otimes (o, \delta) \\
= \; & (1 \land o, \lambda a.( (\lambda b. \emptyset)(a) ( o \cup (\digcup_{b \in A} \{ b \} \delta(b))) \cup i(1) \delta(a))) \\
= \; & (1 \land o, \lambda a.(\emptyset \cup \{ \epsilon \} \delta(a))) \\
= \; &  (o, \delta)
\end{split}\end{equation*}\vspace{3 pt}

\item $(o, \delta) \otimes \mathtt{1} = (o, \delta)$:
\begin{equation*}\begin{split}
& (o, \delta) \otimes \mathtt{1} \\
= \; & (o, \delta) \otimes (1, \lambda a. \emptyset) \\
= \; & (o \land 1, \lambda a.( \delta(a) ( i(1) \cup (\digcup_{b \in A} \{ b \} (\lambda c. \emptyset)(b))) \cup i(o) (\lambda b. \emptyset)(a))) \\
= \; & (o \land 1, \lambda a.( \delta(a) ( i(1) \cup (\digcup_{b \in A} \{ b \} \emptyset)) \cup \emptyset)) \\
= \; & (o \land 1, \lambda a.( \delta(a) ( i(1) \cup \emptyset) \cup \emptyset)) \\
= \; & (o, \delta)
\end{split}\end{equation*}\vspace{3 pt}

\item $((o_1, \delta_1) \otimes (o_2, \delta_2)) \otimes (o_3, \delta_3) = (o_1, \delta_1) \otimes ((o_2, \delta_2) \otimes (o_3, \delta_3))$:
\begin{eqnarray*}
& & ((o_1, \delta_1) \otimes (o_2, \delta_2)) \otimes (o_3, \delta_3) \\
& = & (o_1 \land o_2, \lambda a.( \delta_1(a) ( i(o_2) \cup (\digcup_{b \in A} \{ b \} \delta_2(b))) \cup i(o_1) \delta_2(a))) \otimes (o_3, \delta_3) \\
& = & ((o_1 \land o_2) \land o_3, \lambda a. ( \\
& & \qquad \left[\lambda c.( \delta_1(c) ( i(o_2) \cup (\digcup_{b \in A} \{ b \} \delta_2(b))) \cup i(o_1) \delta_2(c))\right](a) \left[i(o_3) \cup (\digcup_{b \in A} \{ b \} \delta_3(b))\right] \\
& & \qquad \qquad \cup \;i(o_1 \land o_2) \delta_3(a) \\
& & )) \\
& = & ((o_1 \land o_2) \land o_3, \lambda a. (\\
& & \qquad
\left[ \delta_1(a) ( i(o_2) \cup (\digcup_{b \in A} \{ b \} \delta_2(b))) \cup i(o_1) \delta_2(a) \right]
 \left[ i(o_3) \cup (\digcup_{b \in A} \{ b \} \delta_3(b)) \right] \\
& & \qquad \qquad \cup \; i(o_1 \land o_2) \delta_3(a) \\
& & )) \\
& = & ((o_1 \land o_2) \land o_3, \lambda a. (\\
& & \qquad i(o_2 \land o_3) \delta_1(a) \\
& & \qquad \qquad \cup \; i(o_3) \delta_1(a) \left[ \digcup_{b \in A} \{ b \} \delta_2(b) \right] \\
& & \qquad \qquad \cup \; i(o_2) \delta_1(a) \left[ \digcup_{b \in A} \{ b \} \delta_3(b) \right] \\
& & \qquad \qquad \cup \; \delta_1(a) \left[ \digcup_{b \in A} \{ b \} \delta_2(b) \right] \left[ \digcup_{b \in A} \{ b \} \delta_3(b) \right] \\
& & \qquad \qquad \cup \; i(o_1 \land o_3) \delta_2(a) \\
& & \qquad \qquad \cup \; i(o_1)\delta_2(a)\left[ \digcup_{b \in A} \{ b \} \delta_3(b) \right] \\
& & \qquad \qquad \cup \; i(o_1 \land o_2) \delta_3(a) \\
& & )) \\
& = & (o_1 \land (o_2 \land o_3), \lambda a. (\\
& & \qquad \delta_1(a)\left[ i(o_2 \land o_3) \cup \digcup_{b \in A} \{ b \} \left[ 
\delta_2(b)\left[ i(o_3) \cup \digcup_{c \in A} \{ c \} \delta_3(c) \right] \cup i(o_2) \delta_3(b) \right] \right] \\
& & \qquad \qquad \cup \; i(o_3) \left[ \delta_2(a) \left[ i(o_3) \cup \digcup_{b \in A} \{ b \} \delta_3(b) \right] \cup i(o_2)\delta_3(a) \right] \\
& & )) \\
& = & (o_1, \delta_1) \otimes (o_2 \land o_3, \lambda a.( \delta_2(a) ( i(o_3) \cup (\digcup_{b \in A} \{ b \} \delta_3(b))) \cup i(o_2) \delta_3(a))) \\
& = & (o_1, \delta_1) \otimes ((o_2, \delta_2) \otimes (o_3, \delta_3))
\end{eqnarray*}

\item $((o_1, \delta_1) \oplus (o_2, \delta_2)) \otimes (o_3, \delta_3) = ((o_1, \delta_1) \otimes (o_3, \delta_3)) \oplus ((o_2, \delta_2) \otimes (o_3, \delta_3))$:
\begin{eqnarray*}
& & ((o_1, \delta_1) \oplus (o_2, \delta_2)) \otimes (o_3, \delta_3) \\
& = & (o_1 \lor o_2, \lambda a.(\delta_1(a) \cup \delta_2(a))) \otimes (o_3, \delta_3) \\
& = & ((o_1 \lor o_2) \land o_3, \lambda a.( (\lambda b.(\delta_1(b) \cup \delta_2(b)))(a) \left[ i(o_3) \cup (\digcup_{b \in A} \{ b \} \delta_3(b)) \right] \cup i(o_1 \lor o_2) \delta_3(a))) \\
& = & ((o_1 \land o_3) \lor (o_2 \land o_3), \lambda a.( \\
& & \qquad \delta_1(a) \left[ i(o_3) \cup (\digcup_{b \in A} \{ b \} \delta_3(b)) \right] \cup i(o_1) \delta_3(a)\\
& & \qquad \qquad \cup \; \delta_2(a) \left[ i(o_3) \cup (\digcup_{b \in A} \{ b \} \delta_3(b)) \right] \cup i(o_2) \delta_3(a) \\
& & )) \\
& = & ((o_1, \delta_1) \otimes (o_3, \delta_3)) \oplus ((o_2, \delta_2) \otimes (o_3, \delta_3))
\end{eqnarray*}\vspace{-4 pt}

\item $(o_1, \delta_1) \otimes ((o_2, \delta_2) \oplus (o_3, \delta_3)) = ((o_1, \delta_1) \otimes (o_2, \delta_2)) \oplus ((o_1, \delta_1) \otimes (o_3, \delta_3))$:
\begin{eqnarray*}
& & (o_1, \delta_1) \otimes ((o_2, \delta_2) \oplus (o_3, \delta_3)) \\
& = & (o_1, \delta_1) \otimes (o_2 \lor o_3, \lambda a.(\delta_2(a) \cup \delta_3(a))) \\
& = & (o_1 \land (o_2 \lor o_3), \lambda a.( \\
& & \qquad\delta_1(a) ( i(o_2 \lor o_3) \cup (\digcup_{b \in A} \{ b \} (\lambda c.(\delta_2(c) \cup \delta_3(c)))(b))) \cup i(o_1) (\lambda c.(\delta_2(c) \cup \delta_3(c)))(a) \\
& & )) \\
& = & (o_1 \land (o_2 \lor o_3), \lambda a.( \\
& & \qquad\delta_1(a) ( i(o_2 \lor o_3) \cup (\digcup_{b \in A} \{ b \} (\delta_2(b) \cup \delta_3(b)))) \cup i(o_1) (\delta_2(a) \cup \delta_3(a)) \\
& & )) \\
& = & ((o_1 \land o_2) \lor (o_1 \land o_3)), \lambda a.( \\
& & \qquad\delta_1(a) ( i(o_2) \cup (\digcup_{b \in A} \{ b \} \delta_2(b))) \cup i(o_1) \delta_2(a) \\
& & \qquad \qquad \cup \; \delta_1(a) ( i(o_3) \cup (\digcup_{b \in A} \{ b \} \delta_3(b))) \cup i(o_1) \delta_3(a) \\
& & )) \\
& = & ((o_1, \delta_1) \otimes (o_2, \delta_2)) \oplus ((o_1, \delta_1) \otimes (o_3, \delta_3))
\end{eqnarray*}\vspace{-4 pt}

\item $(o, \delta) \otimes \mathtt{0} = \mathtt{0}$:
\begin{eqnarray*}
& & (o, \delta) \otimes \mathtt{0} \\
& = & (o, \delta) \otimes (0, \lambda a. \emptyset) \\
& = & (o \land 0, \lambda a.( \delta(a) ( i(0) \cup (\digcup_{b \in A} \{ b \} (\lambda c.\emptyset)(b))) \cup i(o) (\lambda c. \emptyset)(a))) \\
& = & (0, \lambda a. \emptyset) \\
& = & \mathtt{0}
\end{eqnarray*}\vspace{-4 pt}

\item $\mathtt{0} \otimes (o, \delta) = \mathtt{0}$:
\begin{eqnarray*}
& & \mathtt{0} \otimes (o, \delta) \\
& = & (0, \lambda a. \emptyset) \otimes (o, \delta) \\
& = & (0 \land o, \lambda a. ((\lambda b.\emptyset)(a) (i(o) \cup (\digcup_{b \in A} \{ b \} \delta(b))) \cup i(0) \delta(a))) \\
& = & (0 \land o, \lambda a. (\emptyset (i(o) \cup (\digcup_{b \in A} \{ b \} \delta(b))) \cup \emptyset \delta(a))) \\
& = & (0, \lambda a. \emptyset)
\end{eqnarray*}

\item $(o, \delta) \oplus (o, \delta) = (o, \delta)$:
\[
(o, \delta) \oplus (o, \delta) = (o \lor o, \lambda a.(\delta(a) \cup \delta(a))) = (o, \delta)\eqno{\qEd}
\]
\end{iteMize}

\bibliography{biblio}{}
\bibliographystyle{alpha}

\end{document}